\documentclass[useAMS,usenatbib]{mn2e}
\usepackage{graphicx}
\usepackage{lscape}
\usepackage{times}
\usepackage{amsmath}
\usepackage{amsfonts}
\usepackage{amssymb}
\usepackage{multirow}
\usepackage{longtable}
\usepackage{afterpage}
\usepackage{pdflscape}

\RequirePackage{color}

\title[Fundamental properties of $Kepler$ and $CoRoT$ targets]{
Fundamental properties of $Kepler$ and $CoRoT$ targets: III. Tuning scaling relations using the first 
adiabatic exponent
}
\author[M. Y\i ld\i z, Z. \c{C}elik and C. Kayhan]{M. Y\i ld\i z\thanks{E-mail:
mutlu.yildiz@ege.edu.tr}, Z. \c{C}elik Orhan and C. Kayhan \\ 
$ $\\
Department of Astronomy and Space Sciences, Science Faculty, Ege University, 35100, Bornova, \.Izmir, Turkey\\
}

\begin{document}
\date{Accepted 2013 May 15. Received 2013 April 11; in original form 2013 April 11}

\pagerange{\pageref{firstpage}--\pageref{lastpage}} \pubyear{2013}
\def\braket#1{\left<#1\right>}
\newcommand{\yildiz}{Y\i ld\i z }
\newcommand{\etal}{et al. }
\newcommand{\wrt}{with respect to }
\newcommand{\logg}{\log(g) }
\newcommand{\numino}{\mbox{\ifmmode{\overline{\nu_{\rm min}}}\else$\overline{\nu_{\rm min}}$\fi}}
\newcommand{\numin}{\mbox{\ifmmode{\nu_{\rm min}}\else$\nu_{\rm min}$\fi}}
\newcommand{\teff}{\mbox{\ifmmode{T_{\rm eff}}\else$T_{\rm eff}$\fi}}
\newcommand{\teffsun}{\mbox{\ifmmode{{\rm T}_{\rm eff\sun}}\else${\rm T}_{\rm eff\sun}$\fi}}
\newcommand{\numax}{\mbox{$\nu_{\rm max}$}}
\newcommand{\nuH}{\mbox{\ifmmode{\nu_{\rm minH}}\else$\nu_{\rm minH}$\fi}}
\newcommand{\nuL}{\mbox{\ifmmode{\nu_{\rm minL}}\else$\nu_{\rm minL}$\fi}}
\newcommand{\Dnu}{\mbox{$\Delta \nu$}}
\newcommand{\muHz}{\mbox{$\mu$Hz}}
\newcommand{\kepler}{\mbox{\it{Kepler}}}
\newcommand{\corot}{\mbox{\it{CoRoT}}}
\newcommand{\numaxS}{\mbox{$\nu_{\rm max \sun}$}}
\newcommand{\MS}{{\rm M}\ifmmode_{\sun}\else$_{\sun}$~\fi}
\newcommand{\RS}{{\rm R}\ifmmode_{\sun}\else$_{\sun}$~\fi}
\newcommand{\LS}{{\rm L}\ifmmode_{\sun}\else$_{\sun}$~\fi}
\newcommand{\MSbit}{{\rm M}\ifmmode_{\sun}\else$_{\sun}$\fi}
\newcommand{\RSbit}{{\rm R}\ifmmode_{\sun}\else$_{\sun}$\fi}
\newcommand{\LSbit}{{\rm L}\ifmmode_{\sun}\else$_{\sun}$\fi}
\maketitle
\label{firstpage}
\begin{abstract}
{So called scaling relations have the potential to reveal the mass and radius of solar-like oscillating stars,
based on oscillation frequencies.}
In derivation of these relations, it is assumed that
the first adiabatic exponent at the surface ($\Gamma_{\rm \negthinspace 1s}$) of such stars  is constant. However, 
by constructing interior models for the mass range 0.8-1.6 \MSbit,
we { show} that $\Gamma_{\rm \negthinspace 1s}$ is not constant at stellar surfaces for the effective temperature range {with which} we deal.
Furthermore, the {well-known} relation between large separation and mean density also depends on $\Gamma_{\rm \negthinspace 1s}$.
{Such knowledge is the basis for our aim of modifying scaling relations.}
There are significant differences between masses and radii found from
modified and conventional scaling relations.
{ However, comparison of predictions of these relations with the non-asteroseismic observations of Procyon A
{reveals} that new scaling relations are {effective in determining the} mass and radius of stars.}
In the present study, solar-like oscillation frequencies of 89 target stars (mostly {\it Kepler } and {\it CoRoT})  {were} analysed.
{As well as two new reference frequencies ($\nu_{\rm min1}$ and  $\nu_{\rm min2}$) found in the spacing of solar-like
oscillation frequencies of stellar interior models, we also take into account $\nu_{\rm min0}$.}
In addition to the frequency of maximum amplitude, 
these frequencies have very strong diagnostic potential for determination of
fundamental properties. {The present} study {involves the} application of derived relations from the models to the
solar-like oscillating stars,
and { computes} their effective temperatures using purely asteroseismic methods. 
There are in general very {close} agreements between effective temperatures from asteroseismic and non-asteroseismic (spectral and photometric) methods.
For the Sun and { Procyon A}, for example, the agreement is almost {total}.


%
\end{abstract}


\section{Introduction}
%
%

{Many different types of physical processes occur deep inside stars.}
The standard stellar models (SSMs), however, are constructed by taking into account only {the most basic} of the most essential 
physical principles (such as structure equations, matter-matter and matter-radiation interactions). Therefore, it is {inevitable} that
our SSMs {will be} inadequate in some respects.
{Progress} depends on {finding} discrepancies between SSMs and stars.
Very detailed analysis is required for improvement
in stellar physics, and also for discovering new processes occuring inside stars, {and only} very precise constraints can lead to {the discovery of such processes.
Helioseismology and asteroseismology of solar-like oscillating stars are able to provide such constraints}
(see e.g. Kosovichev 2011, Chaplin \& Miglio 2013, Christensen-Dalsgaard 2002, 2016). 

{
Oscillation frequencies { ($\nu$)} of such stars are now available from the 
$Kepler$ (Borucki et al. 2010) and $CoRoT$ (Baglin et al. 2006) space missions, and from ground-based observations (Bedding et al. 2010, Bedding et al. 2007 and Bazot et al. 2012).
According to scaling relations, stellar mass ($M$) and radius ($R$) can be found from frequency of maximum amplitude ($\nu_{\rm max}$), the large separation between the 
oscillation frequencies ($\Delta \nu$) 
and effective temperature (\teff). $\Delta \nu$ is not constant and {therefore} its mean value ($\braket{\Dnu}$) is used in these computations. Furthermore, 
it has an oscillatory component.  
{We have shown in two recent papers (\yildiz $ $ \etal 2014, hereafter Paper I; \yildiz, \c{C}elik Orhan \& Kayhan 2015, hereafter Paper II) 
that 
there is very strong diagnostic potential in 
the new reference frequencies ($\nu_{\rm min1}$ and  $\nu_{\rm min2}$) at which \Dnu$ $ is minimum. }
}


{Paper I involved the investigations of models constructed using the {\small ANK\.I } code for the solar-like oscillating stars with solar composition, 
revealing for the first time new relations between oscillation  frequencies and fundamental stellar parameters.}
The discovery of new reference frequencies $\nu_{\rm min1}$ and  $\nu_{\rm min2}$ made these relations available.
In order to derive general relations, the effects of metallicity ($Z$) and helium abundance ($Y$) should be clarified.
In Paper II, we {attempted} to generalize the relations for { arbitrary $M$, $R$,  $Z$ and  $Y$ values}.
%
In the present study, we analyse observed oscillation frequencies, {confirm} the relations between
the reference frequencies, and apply the new methods to 
the $Kepler$ and $CoRoT$ target stars. {Their 
\teff, $R$ and $M$ were found using}  asteroseismic parameters. 
Such an application is also very important for testing the scaling relations.


In derivation of the scaling relations used to compute { $M$ and $R$ in terms of \numax, $\braket{\Dnu}$ and $\teff$, }
it is assumed that the first adiabatic exponent ($\Gamma_{\rm \negthinspace 1s}$)  and mean molecular weight ($\mu$) at stellar surface are constant (Brown et al. 1991; Kjeldsen \& Bedding 1995). 
We test {whether} these quantities
are constant {for our purposes and, if not, recommend that the scaling relations should be verified (see Section 3).}
 
The relation { between $\braket{\Dnu}$ and } mean density ($\braket{\rho}$) 
{has been} widely discussed in recent papers. White et al. (2011) state that the $\braket{\Dnu}$-$\braket{\rho}$ relation
depends { on \teff$ $ and} suggest a fitting formula for correction in order to find $\braket{\rho}$  from
the observed values of $\braket{\Dnu}$.
In addition to the $\braket{\Dnu}$-$\braket{\rho}$ relation, Belkacem et al. (2013) also discuss the relation between \numax$ $ and 
acoustic cut-off frequency ($\nu_{\rm ac}$), {claiming} that departure from the observed relation arises from the complexity of 
non-adiabatic processes.
Garc{\'{i}}a Hern{\'{a}}ndez et al. (2015), however,  derive an observational scaling ($\braket{\Dnu}$-$\braket{\rho}$) relation for the $\delta$ Scuti components 
in eclipsing binaries. 
Recently, Sharma et al. (2016) {aimed} to generalize problems pertaining to the scaling relations for \kepler$ $ red giants,
without considering $\Gamma_{\rm \negthinspace 1s}$ as variable.
Our strategy is first to {identify whether} $constants$ are constant, {before attempting to find} parameters for $\braket{\Dnu}$-$\braket{\rho}$ and \numax-$\nu_{\rm ac}$
relations.


For the assessment of the results on fundamental properties of stars, it is {crucial to derive the observed values 
of these parameters 
by alternative direct methods, 
for which, the roles of Sun and { Procyon A} are of key importance.}
From astrometric observations of Procyon by $Hubble$ $Space$ $Telescope$,  {the} mass of its primary component is determined very precisely, $1.478 \pm 0.012$ \MS
(Bond et al. 2015). This data {enables the testing of} new scaling relations (see Section 4).
In addition to interferometrically observed solar-like oscillating stars (Huber et al. 2011b, Baines et al. 2014), 
component stars in eclipsing binaries are benchmark for asteroseismic studies (Gaulme et al. 2013, Rawls et al. 2016),
despite the {complicating factor of} tidally induced oscillations.

This paper is organized as follows. 
{Section 2 is the presentation of the basic asteroseismic and non-asteroseismic properties of the target stars compiled from literature.}
In this section, we also compare $\nu_{\rm min1}$ and $\nu_{\rm min2}$ of the models with their observational counterparts.
Section 3 is devoted to {\small MESA} { (Paxton et al. 2011)} models and role of $\Gamma_{\rm \negthinspace 1s}$ in new scaling relations. 
In this section, we also develop new expressions for effective temperature by using  oscillation frequencies.
In Section 4, {the results based on asteroseismic methods are presented and compared} with results obtained by conventional
methods.
Finally, in Section 5, {conclusions are drawn}.

\section{Asteroseismic and non-asteroseismic properties of \kepler$ $ and \corot$ $  solar-like oscillating stars}



The basic data of {certain} \kepler$ $ (79 stars) and \corot$ $ (7 stars) target stars,  compiled from the literature, are listed 
in Table A1. Oscillation frequencies of three stars ({Procyon A}, HD 2151 and HD 146233) {were} obtained from ground-based observations (Bedding et al. 2010, Bedding et al. 2007 and Bazot et al. 2012, respectively).
These stars are also listed in this table, {with data for } the Sun for comparison.
For {most stars, we provide} $B-V$ and $V-K$ colours (SIMBAD database) from photometric, and surface gravity ($\logg$), effective temperature 
($T_{\rm eS}$) and metallicity ([Fe/H]) from spectroscopic observations. 

{Observational
oscillation frequencies are obtained from the \kepler$ $ and  \corot$ $ light curves, and from the radial velocity curves. For {most stars},
frequencies ($\nu_{nl}$) of modes with low degrees ($l$) and high order ($n$) are available,
{allowing the computation of} $\numax$, ${\Dnu}$ and small separation between oscillation frequencies 
(${\delta \nu_{02}=\nu_{n0}-\nu_{n-1,2}}$). Mean values of ${\Dnu}$ ($\braket{\Dnu}$) and ${\delta \nu_{02}}$ ($\braket{\delta \nu_{02}}$) are used in scaling relations.
Furthermore, {two minima are seen in the \Dnu-$\nu$ graph of the majority of stars.}
For a few stars, there are more than two minima, {allowing the consideration of 
the effect of He {\scriptsize II} ionization zone on the oscillation frequencies, and the assessment of} their diagnostic potential if any. 
{High frequency minima (minH) and low frequency minima (minL) are refered to as  $\nuH$ and $\nuL$, respectively. }
  
These stars are {plotted on} a $\log(g'_{\rm sca})$-$\log(T_{\rm eS})$ diagram in Fig. 1. $g'_{\rm sca}$ is computed from 
conventional scaling relation ($g'_{\rm sca}/g_{\sun}=\numax/\numax_{\sun}(T_{\rm eS}/\teff_{\sun})^{0.5}$).
Uncertainties in $g'_{\rm sca}$ are computed from uncertainties in $\numax$ and $T_{\rm eS}$ in the standard way:
$\Delta g'_{\rm sca}/g'_{\rm sca}=\Delta \numax/\numax+0.5 \Delta T_{\rm eS}/T_{\rm eS}$.
Also seen in Fig. 1 are zero-age main-sequence (ZAMS) and terminal-age main-sequence (TAMS) lines taken from \yildiz (2015). Nearly half {are} 
main-sequence (MS) stars {while the other half are evolved as far as the red giant phase}.
$T_{\rm eS}$ of the hottest and coolest stars are about 6630 K (KIC 11081729) and 4550 K (KIC 8219268), respectively.
$T_{\rm eS}$ of the  coolest MS star (KIC 11772920) is 5209 K. Therefore, the effective temperature range for the models {that are  used}
to derive relations between asteroseismic and non-asteroseismic quantities is {set} as 5200-6650 K (see Section 3).
\begin{figure}
\includegraphics[width=101mm,angle=0]{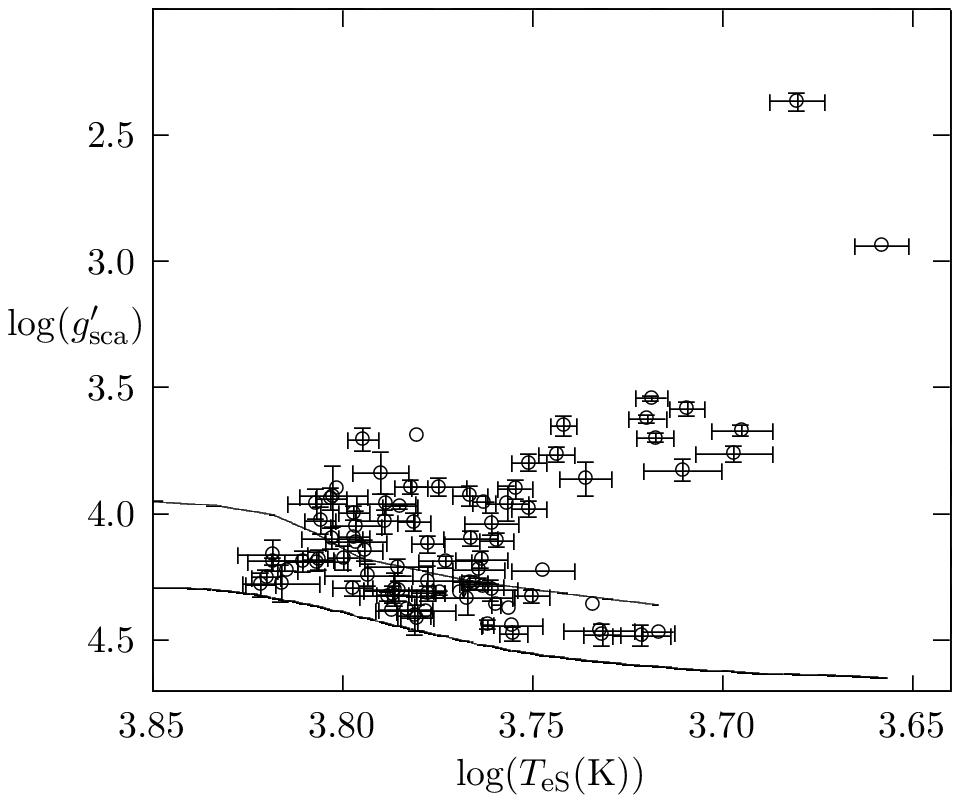}
\caption{  $\log (g'_{\rm sca})$ of the target stars, computed from $\numax T_{\rm eS}^{0.5}$, is plotted \wrt $\log (T_{\rm eS})$. The thin and thick solid lines represent ZAMS and TAMS lines
taken from \yildiz $ $ (2015),
respectively.
}
\end{figure}

{The colours} $B-V$ and $V-K$ derived from atmospheric models (Lejeune, Cuisinier \& Buser 1998) {were fitted} to the observed colours, 
{in order to} find the effective temperatures $T_{\rm eBV}$ and $T_{\rm eVK}$, respectively. In Fig. 2, these effective temperatures 
are plotted \wrt $T_{\rm eS}$.
The three effective temperatures are in good agreement. 
{Only in the case of two stars, namely  KIC 10920273 and HD 146233, there is a large difference between $T_{\rm eVK}$ and $T_{\rm eS}$.
For most stars}, the difference between $T_{\rm eVK}$ and $T_{\rm eS}$, $\delta T_{\rm eVK}=T_{\rm eVK}-T_{\rm eS}$, is as
$-100$ K $<\delta T_{\rm eVK}<100 $ K; 
{few} are out of this range.
{The results for $T_{\rm eBV}$, $-150$ K $<\delta T_{\rm eBV}<150 $ K, are similar, showing} that 
$T_{\rm eVK}$ is in  better agreement with $T_{\rm eS}$ than $T_{\rm eBV}$.
\begin{figure}
\includegraphics[width=101mm,angle=0]{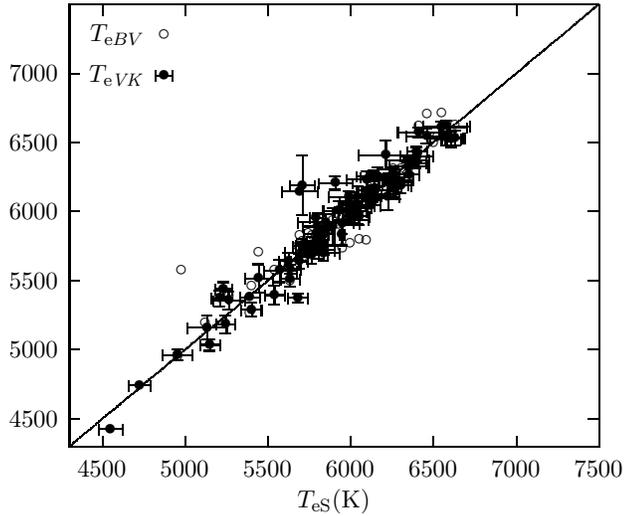}
\caption{Effective temperatures { of the target stars} derived from fitting model colours to the observed colours $B-V$ ($T_{\rm eBV}$) and $V-K$  ($T_{\rm eVK}$) are plotted \wrt $T_{\rm eS}$. 
}
\end{figure}


We have already confirmed { the presence of } $\nu_{\rm min1}$ and $\nu_{\rm min2}$ in the observed oscillation frequencies of the Sun
(BiSON data; Chaplin et al. 1999). 
{It was unexpected to find two such minima 
in the oscillation frequencies
of most} of $Kepler$ and $CoRoT$ target stars. 
As an example \Dnu-$n$ diagram of 16 Cyg A (Metcalfe et al. 2012) is plotted for degrees 
$l=0$ and $1$ in Fig. 3. From data of both degrees, {it was seen} that minL corresponds to the 
mode with order $n=16$. The order of the minima with high frequency for $l=0$ is 21, and 20 for $l=1$. 

{In Fig. 4, $\nuH$ and $\nuL$  from the observed oscillation frequencies of the target stars are plotted \wrt \numax}. 
The most striking result is that for {most,}
$\nuH$ is greater than \numax$ $ while $\nuL$ is lower than \numax. For the {\small ANK\.I } models (Paper I and II), the situation 
{was found to be }
different: $\nu_{\rm min1}$ is less than \numax$ $
if $M< 1.2 \MS$, otherwise $\nu_{\rm min1} >$ \numax. This inconsistency may arise from mismatch of the minima, 
i.e. minH may not be min1, and
similarly min2 may not be minL. Another {possibility} is that the ordering of these frequencies according to their values 
is model dependent (see below). 
\begin{figure}
\includegraphics[width=101mm,angle=0]{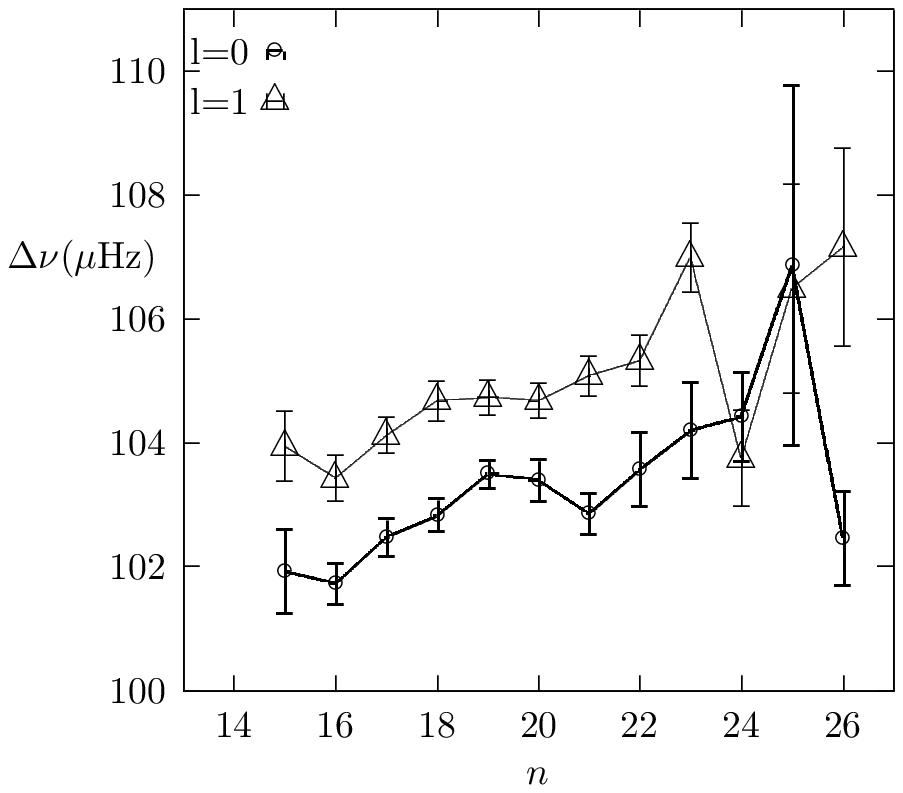}
\caption{  $\Delta \nu$ is plotted with respect to order $n$ for the observed oscillation frequencies of 16 Cyg A 
(Metcalfe et al. 2012). The circle and triangle show the modes with $l=0$ and $l=1$, respectively.
\Dnu$ $ for $l=1$ is shifted up 1.25 \muHz$ $ for a clear appearance. 
Two minima appear for $l=0$; one is about $n=16$ and the other is about $n=21$. Notice that 
there are large scattering for the modes with $n>23$. A similar scattering is seen in the early
helioseismic data (see e.g. Grec, Fossat \& Pomerantz 1983).
}
\end{figure}

\begin{figure}
\includegraphics[width=101mm,angle=0]{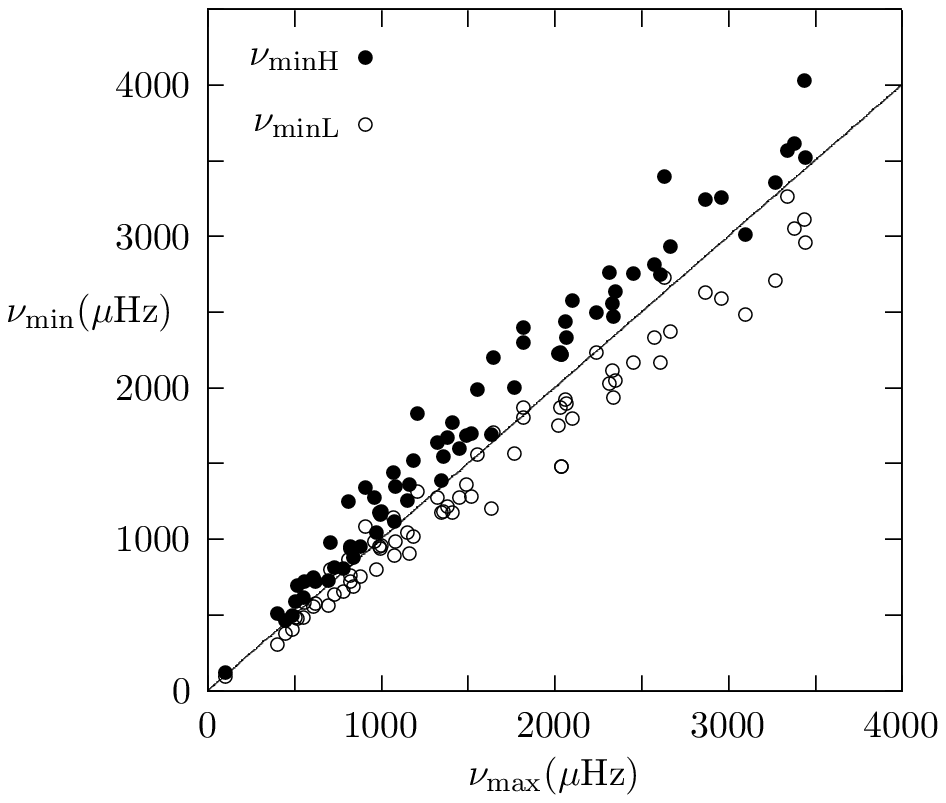}
\caption{ { $\nu_{\rm minH}$ (filled circles) and  $\nu_{\rm minL}$ (circles) of the observed frequencies are plotted 
with respect to $\nu_{\rm max}$.}
}
\end{figure}

In \Dnu-$\nu$ graph for the {\small ANK\.I } models (see figure 3 of Paper I), the depth around $\nu_{\rm min1}$ is deeper than
that of around $\nu_{\rm min2}$ for 1.0 \MS$ $ {models, and} comparable to that of $\nu_{\rm min2}$ for 1.2 \MS$ $ models.
We notice {considerable difference in the depths of minH and minL in \Dnu-$\nu$ graph for observed oscillation frequencies.} 
The depth of minH is in general {much} shallower than that of minL. 
In figure 3 of Paper I, very clear shallow minima are seen in the high frequency ($\nu > \nu_{\rm min1}$) range for the 
models with 1.1 and 1.2 \MSbit.

Although the solar models constructed by using the {\small ANK\.I } code are in good agreement with the helioseismic data of the Sun,
there are significant differences between asteroseismic data of the solar-like oscillating stars and models, at least for $\nu_{\rm min1}$, $\nu_{\rm min2}$ and \numax.
Therefore, we construct new models {using a different code, to make a comparison between  their asteroseismic properties and} the observed ones.

\section{Properties of {\small MESA} models and role of $\Gamma_{\rm \negthinspace 1s}$ in scaling relations}
Our results presented in Paper I and II are based on the {\small ANK\.I } models.
It {is important to test whether} these results are code dependent. 
Furthermore, as emphasized above, 
 {there are large differences between models and the observed oscillation frequencies in terms of the
relations between \numax$ $ and frequencies 
of minima}.
Therefore, 
we construct models by using the  {\small MESA} evolution code (Paxton et al. 2011, 2013).
As in the case of the {\small ANK\.I } models, we obtain solar values by fitting interior model to the Sun, 
{
then use these values to}   construct stellar models. 
The solar values for the {\small MESA} code for initial hydrogen abundance ($X$), $Z$ and the mixing-length parameter ($\alpha$) are $X=0.70358$, $Z=0.0172$ and $\alpha=2.175$.
As in the case of the {\small ANK\.I } models, we also construct models by using {\small MESA} 
for the  mass range 0.8 \MS $\le M \le$ 1.6 \MS and compute { their} adiabatic oscillation frequencies when the
central hydrogen abundance { ($X_{\rm c}$)} is  {approximately} $X_{\rm c}=0.7$, $0.53$, $0.35$ and $0.17$.

In  {the } construction of interior models with {\small MESA}, standard mixing-length theory (B{\"{o}}hm-Vitense 1958)  {was} derived for convection treatment,
whereas the effects of convective overshooting  {were} not considered.
 {{\small MESA} {\it EOS} tables were selected for equation-of-state, and 
OPAL opacity tables (Iglesias \& Rogers 1993, 1996) were used } in the high temperature region 
supplemented by the low-temperature tables of Ferguson et al. (2005).
Nuclear reaction rates {were} taken from Angulo et al. (1999) with significant updates (Kunz et al. 2002; Cyburt et al. 2010).
The element diffusion is used in {\small MESA} default option (see in detail Paxton et al. 2011) for {the solar model only}.
As stellar atmosphere, we choose \texttt{simple\_photosphere} in our models.
Adiabatic oscillation frequencies {were} computed using {\small ADIPLS} oscillation package (Christensen-Dalsgaard 2008) in the {\small MESA} module.

{It is important first to clarify which of the minima found} in observed oscillation frequencies ($\nu_{\rm minH}$ and $\nu_{\rm minL}$) matches
min1 or min2. In Fig. 5, both model (min1  and min2) and observed (minH  and minL) frequencies of minima 
are plotted with respect to $\nu_{\rm max}$. 
We notice that {while $\nu_{\rm min1}$ and $\nu_{\rm min2}$ do not exactly match
$\nu_{\rm minH}$ and $\nu_{\rm minL}$,} $\nu_{\rm minL}$ and $\nu_{\rm min1}$ are in good agreement for {the majority} of the data. 
{Revisiting the  \Dnu-$\nu$ graph of some models, we find 
the frequencies of shallower minima (see e.g. fig. 3 in Paper I)}. Frequencies ($\nu_{\rm min0}$) of
these minima (called as min0, in accordance with min1 and min2) are also plotted in Fig. 5. We confirm that $\nu_{\rm minH}$ corresponds 
$\nu_{\rm min0}$, for most of the stars. 
\begin{figure}
\includegraphics[width=101mm,angle=0]{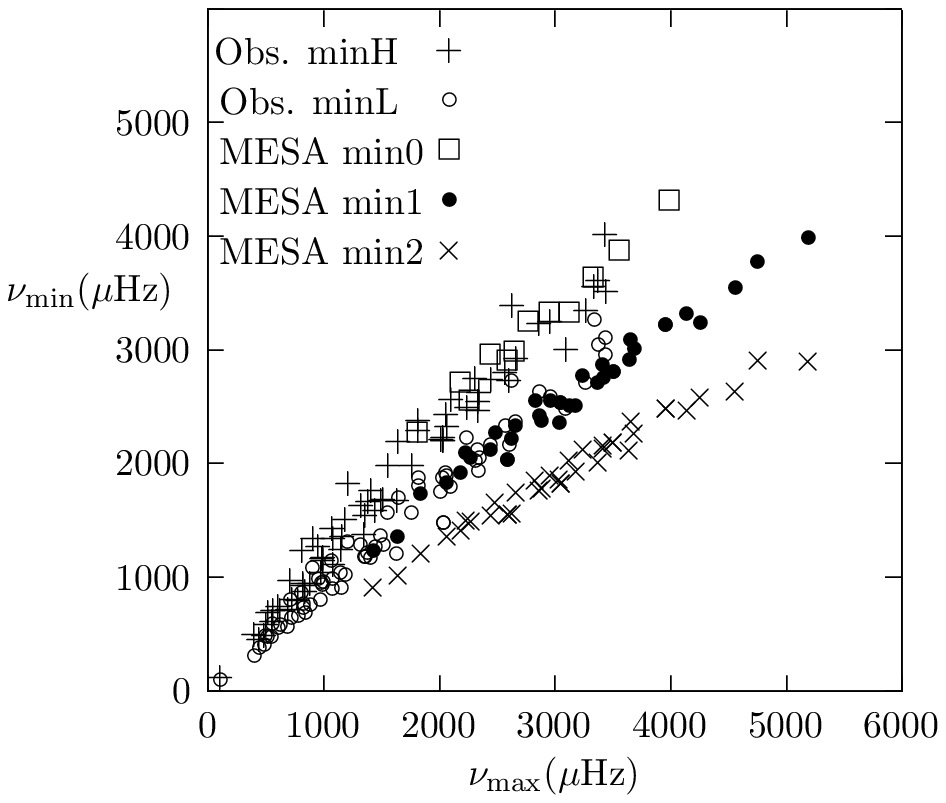}
\caption{Comparison of model and observational frequencies of minima in \Dnu-$\nu$ graph.
{ $+$ and circle show observed oscillation frequencies of minH and minL, respectively. 
Square, filled circle and cross are for min0, min1 and min2 of model frequencies, {respectively}.}
}
\end{figure}


%
%
%
%
%
%
%

\subsection{Why $\Gamma_{\rm \negthinspace 1s}$ is important in scaling relations?}
In asteroseismic studies, \numax$ $ (Brown et al. 1991) is taken as 
\begin{equation}
\nu_{\rm max}\propto \frac{g}{\sqrt{\teff}}.
\end{equation}
{ \numax$ $ is assumed to be proportional to 
$\nu_{\rm ac}$ (Lamb 1909; Balmforth \& Gough 1990), given as}
\begin{equation}
\nu_{\rm ac}\sim \frac{c}{H}\left(1-\frac{dH}{dr}\right),
\end{equation}
where $c$ and $H$ are sound speed and pressure scale height at the
stellar surface, respectively. $dH/dr$, { gradient of $H$,} has order about $10^{-4}$ and henceforth is negligibly small.
If we insert the usual expressions for $c$ and $H$ in equation ({2}), we obtain
\begin{equation}
\nu_{\rm ac}= \frac{g}{\sqrt{\mathfrak{R}\teff}} \sqrt{\Gamma_{\rm \negthinspace 1s} \mu}.
\end{equation}
where $\mathfrak{R}$ is the multiplication of Boltzmann constant and Avagadro's number. 
It is {clear} that equation ({1}) is valid { if $\Gamma_{\rm \negthinspace 1s}$ and $ \mu$ are constant for the 
solar-like oscillating stars}. We {confirm below that this is the case}.

\begin{figure}
\includegraphics[width=101mm,angle=0]{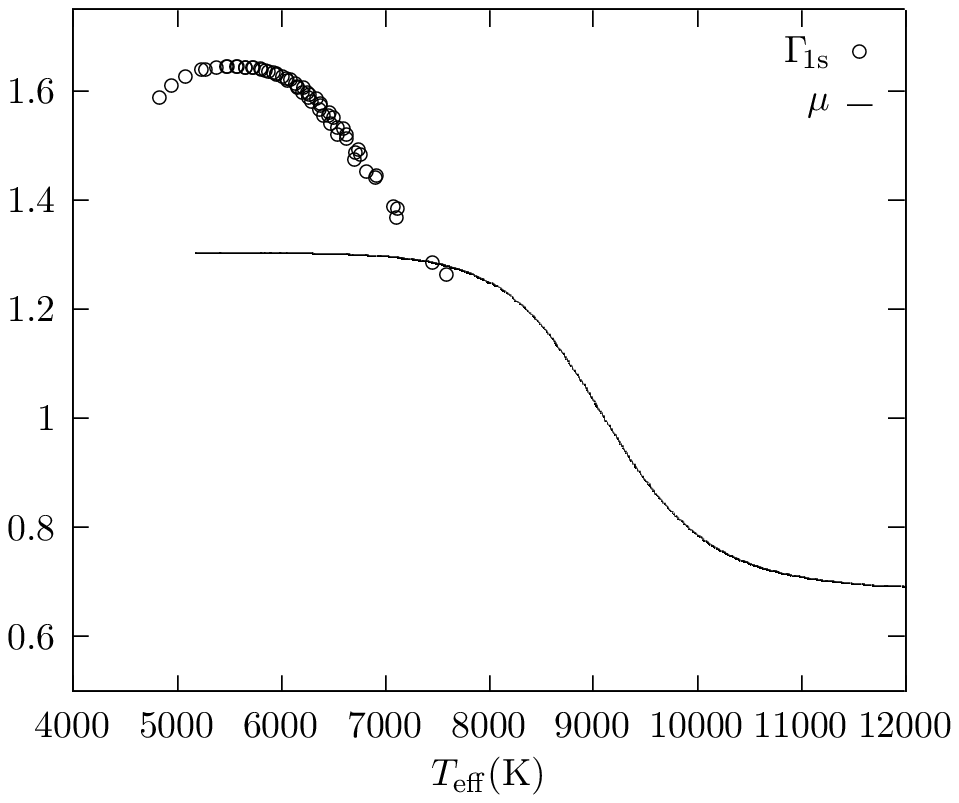}
\caption{$\Gamma_{\rm \negthinspace 1s}$ { (circle) and $\mu$ (line) } at the surface of interior models 
\wrt \teff.
}
\end{figure}
In Fig. 6, $\Gamma_{\rm \negthinspace 1s}$ and $ \mu$ at the surface of interior models 
are plotted with respect to \teff. For the cool stars, both hydrogen and helium 
{ are}  neutral and $\mu$ is about 1.3. { $\mu$  decreases}  as hydrogen and helium are ionized. 
For the solar-like oscillating stars 
($\teff <6900$ K), $ \mu$ is constant in great extent. However, this is not the
case for $\Gamma_{\rm \negthinspace 1s}$. Its maximum value is about 1.64 at about $\teff=5500$ K and it decreases 
{for higher or lower $\teff$ values.}
The value of $\Gamma_{\rm \negthinspace 1s}$ for the hottest solar-like oscillating stars
is 1.25, nearly 30 per cent {below} the maximum value. Therefore, 
{to improve the accuracy of the scaling relation, it is important to take into account variation of $\Gamma_{\rm \negthinspace 1s}$.}

%

$\Gamma_{\rm \negthinspace 1s}$ is a function of \teff. We fit a parabolic function for $1/\Gamma_{\rm \negthinspace 1s}$ to the data shown in Fig. 6. The derived function is as
\begin{equation}
\frac{1}{\Gamma_{\rm \negthinspace 1s}}= 1.6 \left(\frac{\teff}{\teff_\odot}-0.96\right)^2+0.607.
\end{equation}
{Equation ({4}) is very effective in representing $\Gamma_{\rm \negthinspace 1s}$ for the cool stars in the (\teff) range under consideration in this study.
For red giants, for example,} there
might be a deviation.
If we adopt that $\Gamma_{\rm \negthinspace 1s}$ is not constant, we obtain new scaling relations for stellar mass { ($M''_{\rm sca}$)} as 
\begin{equation}
\frac{M''_{\rm sca}}{M_{\odot}}=\frac{(\numax/\nu_{\rm max\odot})^3}{(\braket{\Delta \nu}/\braket{\Delta \nu_\odot})^4}\left( \frac{T_{\rm eff}}{T_{\rm eff\odot}}\frac{\Gamma_{\rm \negthinspace 1s\odot}}{\Gamma_{\rm \negthinspace 1s}}\right)^{3/2}
\end{equation}
and for radius { ($R''_{\rm sca}$)} as
\begin{equation}
\frac{R''_{\rm sca}}{R_{\odot}}=\frac{(\numax/\nu_{\rm max\odot})}{(\braket{\Delta \nu}/\braket{\Delta \nu_\odot})^2}\left( \frac{T_{\rm eff}}{T_{\rm eff\odot}}\frac{\Gamma_{\rm \negthinspace 1s\odot}}{\Gamma_{\rm \negthinspace 1s}}\right)^{1/2}.
\end{equation}
In Fig. 7, { $M''_{\rm sca}$ and} mass from conventional scaling relation ($M'_{\rm sca}$) are plotted \wrt {the model} mass ($M_{\rm mod}$) in solar units. 
{Despite the substantial difference between $M'_{\rm sca}$ and $M_{\rm mod}$ in particular for the models with $M > 1.2$ \MSbit,
the inclusion of $\Gamma_{\rm \negthinspace 1s}$ has the effect of increasing the difference, contrary to expectations.}

\begin{figure}
\includegraphics[width=101mm,angle=0]{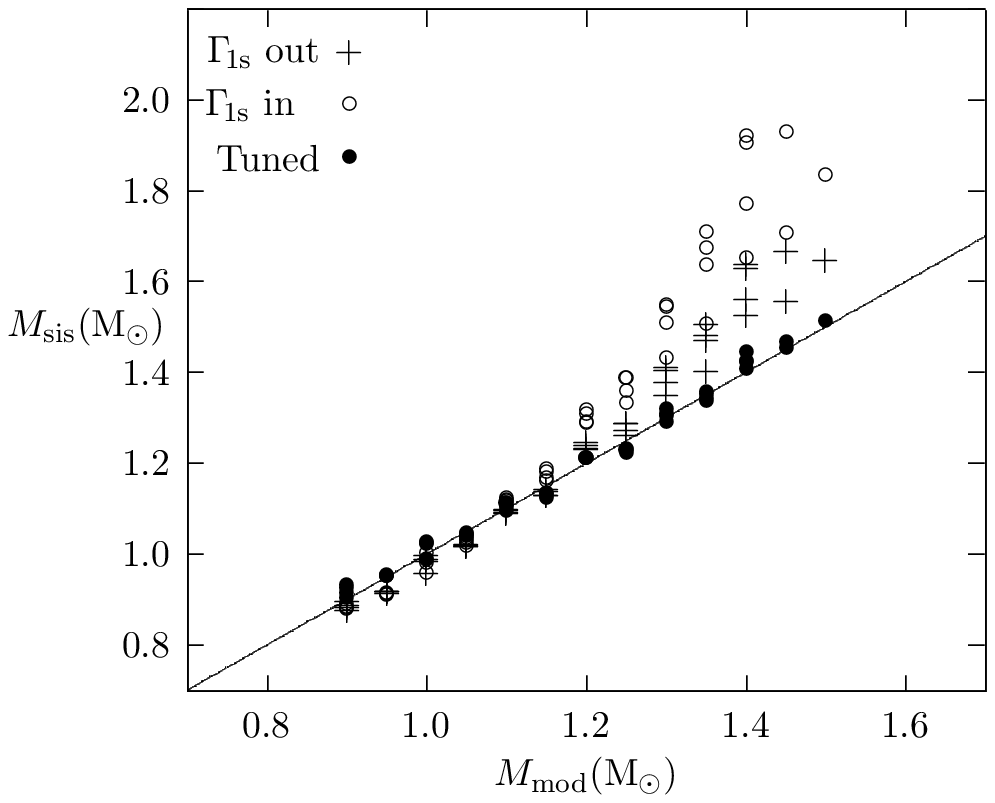}
\caption{Comparison of asteroseismic masses computed from the customary scaling relation ($+$) and the new scaling relation (equation { 5}) with $\Gamma_{\rm \negthinspace 1s}$ 
(circle).  $M_{\rm sca}$ is plotted with respect to model mass in units of solar mass. For the tuned scaling relation (filled circle), see below (equation { 9}). 
}
\end{figure}
We also test the scaling relation for stellar radius in Fig. 8,  
in { which $R''_{\rm sca}$ and} radius from conventional scaling relation ($R'_{\rm sca}$) are  plotted \wrt $R_{\rm mod}$ in solar units. 
\begin{figure}
\includegraphics[width=101mm,angle=0]{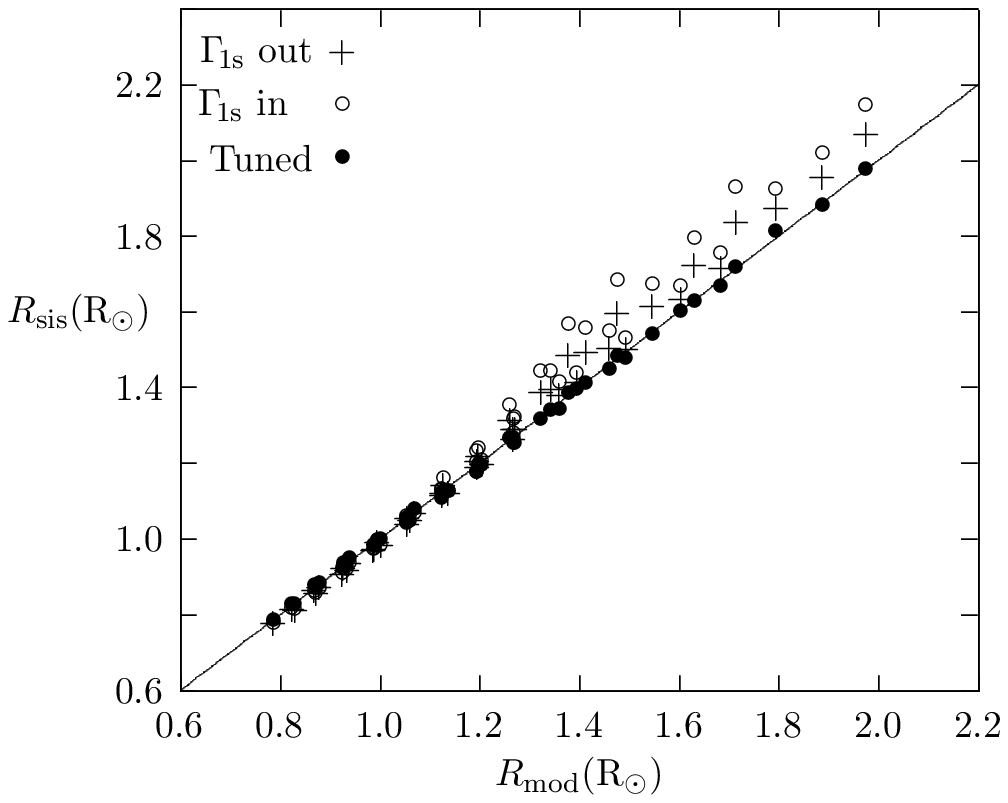}
\caption{Comparison of asteroseismic radii computed from the customary scaling relation ($+$) { and} the new scaling relation (equation { 6}) with $\Gamma_{\rm \negthinspace 1s}$ 
(circle).  $R_{\rm sca}$ is plotted with respect to model radius in units of solar radius.
For the tuned scaling relation (filled circle) for radius, see below (equation { 10}).
}
\end{figure}
There is a significant difference between $R'_{\rm sca}$ and $R_{\rm mod}$ if $R > 1.2$ \RS, {and the greater} difference appears
between $R''_{\rm sca}$ and $R_{\rm mod}$.
This implies that {the inclusion} of $\Gamma_{\rm \negthinspace 1s}$ in equation ({3}) {itself  
does not result in improvement in scaling relations,
highlighting the need for a much more general approach (see below)}.

$\Gamma_{\rm \negthinspace 1s\odot}$, { the solar value of $\Gamma_{\rm \negthinspace 1s}$}, is taken as 1.639. The other solar quantities used in equations (4)-(6) are given at the end of Table A1. 

In scaling relations, we use $T_{\rm eS}$ in place of \teff.  
 {In our analysis,  we use $T_{\rm eVK}$ for those seven stars for which no $T_{\rm eS}$ is available in the literature.}

\subsection{Tuning the scaling relations}
It is reported in some studies (see e.g. White et al. 2011 and Sharma et al. 2016) that the relation between 
$\braket{\Dnu} $ { and $\braket{\rho}^{1/2}$}  deviates from a linear relation. In Fig. 9, { the ratio of}  
$\braket{\Dnu}/\braket{\rho}^{1/2}$ is plotted with respect to $\Gamma_{\rm \negthinspace 1s}$ in solar units. 
There is a very clear 
linear relation between $\braket{\Dnu}/\braket{\rho}^{1/2}$ and  $\Gamma_{\rm \negthinspace 1s}$. The fitting line is found as 
\begin{equation}
\frac{\braket{\Dnu}/\braket{\Dnu_{\sun}}}{(\braket{\rho}/\braket{\rho_{\sun}})^{1/2}}=f_{\Delta \nu}=0.430\frac{\Gamma_{\rm \negthinspace 1s}}{\Gamma_{\rm \negthinspace 1s\odot}}+0.570,
\end{equation}
{  where $f_{\Delta \nu}$ is defined as { the ratio} $\braket{\Dnu}/\braket{\rho}^{1/2}$ in solar units.} 
The range of $f_{\Delta \nu}$ is about [0.95,1.01]. 
In derivation of equation ({7}), we adopt $\braket{\Dnu_\odot}=136$ \muHz. { {Thus}, we obtain its solar value as unity: $f_{\Delta \nu \odot}=1$.}
There are two important indicators for $\braket{\Dnu_\odot}=136$ \muHz: i) The solar oscillation frequencies {in 
Broomhall et al. (2009) yield the same value}; 
ii) the maximum value of \Dnu$ $ in between min1 and min2 is very close to $136 $ \muHz.  
The latter point is important because this mode is 
{the least effected, if not completely unaffected, by the He {\scriptsize II} ionization zone.}
\begin{figure}
\includegraphics[width=91mm,angle=0]{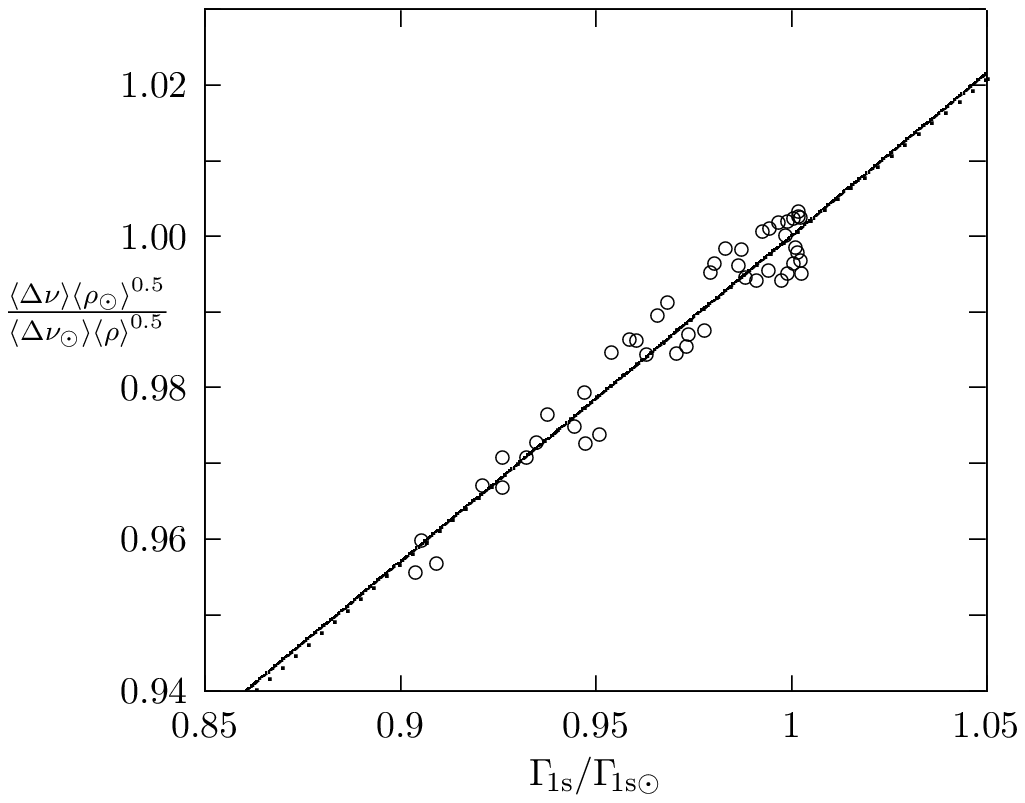}
\caption{
$\braket{\Dnu}/\braket{\rho}^{0.5}$ (in solar units) is plotted \wrt $\Gamma_{\rm \negthinspace 1s}$. 
The solid line is the fitted line $f_{\Delta \nu}=0.43\frac{\Gamma_{\rm \negthinspace 1s}}{\Gamma_{\rm \negthinspace 1s\odot}}+0.57$.
The fitted curve { (dotted line)} is $\left(\frac{\Gamma_{\rm \negthinspace 1s}}{\Gamma_{\rm \negthinspace 1s\odot}}\right)^{0.42}$. These two 
functions are equivalent to each other.
}
\end{figure}

{ 
{Understanding of the} underlying physics of $\Gamma_{\rm \negthinspace 1s}$ dependence of $\braket{\Dnu} / \braket{\rho}^{0.5}$ ratio
{can be achieved by comparing}  
the solar model with a modified solar model (MSM).  Suppose that $\Gamma_{\rm \negthinspace 1s}$ in the most outer region of MSM is lower 
than $\Gamma_{\rm \negthinspace 1s\odot}$, otherwise identical to the solar model.
In the outer region of MSM, sound {travels more slowly than in} the solar model. 
{As speed is decreased in a part of the model, frequencies of all the modes decrease in
accordence with the dispersion relation for sound waves.} 
Radial nodes of MSM come to closer in the modified outer region and move 
away from each other in the unchanged interior regions.  Although these two models have the same mean density, their 
oscillation frequencies are different.  Oscillation frequency of MSM for a given mode ($\nu'_{nl}$) is always less than that of the solar model 
($\nu_{nl\odot}$): $\nu'_{nl}=q \nu_{nl\odot}$, where
$q$ is less than 1, depending on ratio of modified $\Gamma_{\rm \negthinspace 1s}$ to $\Gamma_{\rm \negthinspace 1s\odot}$. Then,
the large separation of MSM ($\Dnu'$) can be written as
$$\Dnu'=\nu'_{nl}-\nu'_{n-1,l}= q(\nu_{nl\odot}-\nu_{n-1,l\odot}) =q \Dnu_\odot.$$
This states that the ratio of $\Dnu'/\Dnu_\odot$ 
for the models with the same density is a function of the ratio 
($\Gamma_{\rm \negthinspace 1s}/ \Gamma_{\rm \negthinspace 1s \odot}$): 
$q=q(\Gamma_{\rm \negthinspace 1s}/ \Gamma_{\rm \negthinspace 1s \odot})$.
}


We know that 
$\nu_{\rm ac}$ is also a function of $\Gamma_{\rm \negthinspace 1s}$. However,
there is another underlying assumption for derivation of the scaling relations, which states 
that $\nu_{\rm ac}/\numax $ is constant. However,
it seems reasonable to assume that 
\begin{equation}
\numax= f_{\nu}\nu_{\rm ac},
\end{equation}
{ where $f_{\nu}$ is a parameter to be determined.} 
Then, we derive new scaling relations for mass { ($M_{\rm sca}$)} and radius { ($R_{\rm sca}$)} as
\begin{equation}
\frac{M_{\rm sca}}{M_{\odot}}=\frac{(\numax/\nu_{\rm max\odot})^3}{(\braket{\Delta \nu}/\braket{\Delta \nu_\odot})^4}\left( \frac{T_{\rm eff}}{T_{\rm eff\odot}}\frac{\Gamma_{\rm \negthinspace 1s\odot}}{\Gamma_{\rm \negthinspace 1s}}\right)^{3/2}
\frac{f_{\Delta \nu}^4}{f_{\nu}^3}
\end{equation}
and 
\begin{equation}
\frac{R_{\rm sca}}{R_{\odot}}=\frac{(\numax/\nu_{\rm max\odot})}{(\braket{\Delta \nu}/\braket{\Delta \nu_\odot})^2}\left( \frac{T_{\rm eff}}{T_{\rm eff\odot}}\frac{\Gamma_{\rm \negthinspace 1s\odot}}{\Gamma_{\rm \negthinspace 1s}}\right)^{1/2}
\frac{f_{\Delta \nu}^2}{f_{\nu}},
\end{equation}
respectively. 
If we plot $f_\nu$ in equation ({9}) by taking $M_{\rm sca}=M_{\rm mod}$, we find a linear relation between $f_\nu$ and ${\Gamma_{\rm \negthinspace 1s\odot}}/{\Gamma_{\rm \negthinspace 1s}}$:
\begin{equation}
f_{\nu}=0.470\frac{\Gamma_{\rm \negthinspace 1s\odot}}{\Gamma_{\rm \negthinspace 1s}}+0.530.
\end{equation}
If we plot $f_\nu$ in equation ({10}) by taking $R_{\rm sca}=R_{\rm mod}$, we find a very similar
$f_{\nu}$: $f_{\nu}=0.456{\Gamma_{\rm \negthinspace 1s\odot}}/{\Gamma_{\rm \negthinspace 1s}}+0.543.$
In our computations, equation ({11}) {is used} for $f_\nu$.
The maximum difference between { $R_{\rm sca}$ (equation 10) and $R_{\rm mod}$} is about 1 per cent
if $f_{\nu} $ {is taken} as in equation ({11}).
The maximum difference between { $M_{\rm sca}$ (equation 9) and $M_{\rm mod}$}
is about 2 per cent.
The range of $f_{\nu}$ is about [0.94,1.0].

{This indicates that $\Gamma_{\rm \negthinspace 1s} $ is a hitherto neglected, but key factor in 
determining relations between asteroseismic and non-asteroseismic quantities.}
The parameters $f_{\Delta \nu}$ and $f_\nu$, appearing in new scaling relations, are plotted in Fig. 10 \wrt \teff.
We insert { expression (4) for $\Gamma_{\rm \negthinspace 1s}$ in equations (7) and (11)}. This gives expressions for $f_{\Delta \nu}$ and $f_\nu$
as functions of \teff. 
We notice that
$\Gamma_{\rm \negthinspace 1s}$, $f_{\Delta \nu}$ and $f_\nu$ are {approximately} unity for the range 5200-5800 K. 
Outside this range, {all}
deviate from {unity, making} conventional scaling relations much more uncertain.
\begin{figure}
\includegraphics[width=91mm,angle=0]{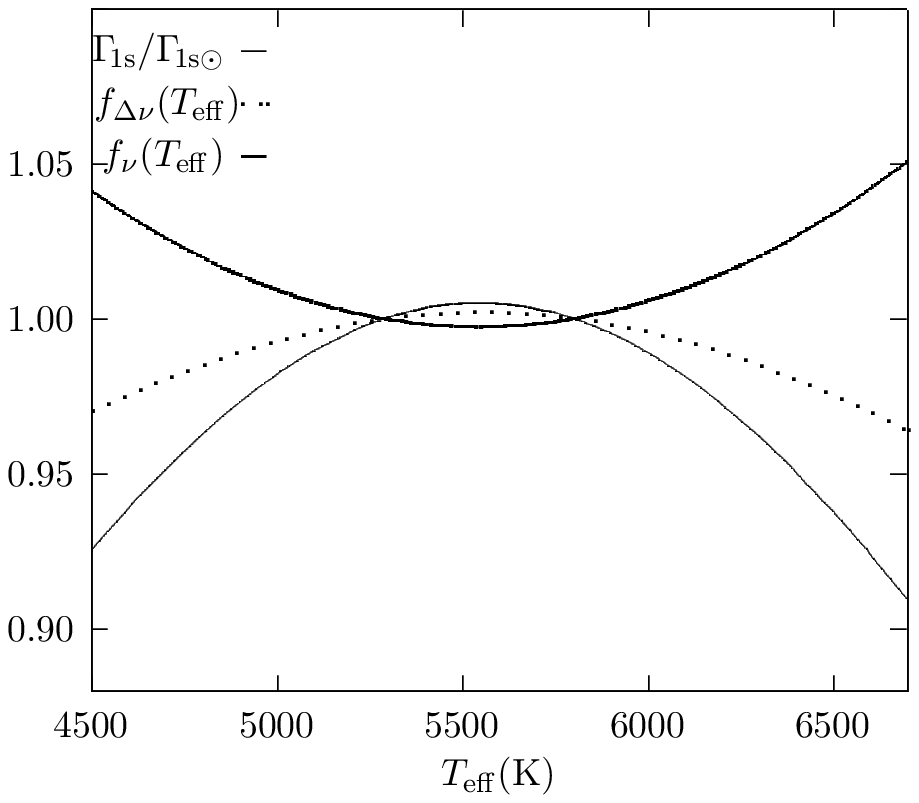}
\caption{$\Gamma_{\rm \negthinspace 1s}/\Gamma_{\rm \negthinspace 1s\odot}$ (thin solid line), $f_{\Delta \nu}$ (dotted line) and $f_\nu$ (thick solid line) are plotted \wrt \teff.
Around $\teff=5500$ K,  they all are about unity. 
}
\end{figure}

$M_{\rm sca}$ and $R_{\rm sca}$ computed from { equations (9) and (10)} are also plotted in Figs. (7) and (8), respectively.
Both of $M_{\rm sca}$ and $R_{\rm sca}$ are in very good agreement with model values.

Uncertainty in $M_{\rm sca}$ can be computed from uncertainties in \numax$ $, \Dnu$ $ and \teff$ $:
\begin{equation}
\frac{\Delta M_{\rm sca}}{M_{\rm sca}}=
3\frac{\Delta \numax}{\numax} 
+4\frac{\Delta \braket{\Delta \nu}}{\braket{\Delta \nu}}
+1.5\frac{\Delta {\teff}}{\teff}.
\end{equation} 
Similarly, uncertainty in radius can be obtained from
\begin{equation}
\frac{\Delta R_{\rm sca}}{R_{\rm sca}}=
\frac{\Delta \numax}{\numax} 
+2\frac{\Delta \braket{\Delta \nu}}{\braket{\Delta \nu}}
+0.5\frac{\Delta {\teff}}{\teff}.
\end{equation}
$\Delta M_{\rm sca}$ and $\Delta R_{\rm sca}$  are computed for the target stars are given in the second row 
of the two lines for each star in Table A1.


\subsection{\teff$ $ from model frequencies}
{In Paper I, for the first time, {\small ANK\.I} models were used to derive a relation between \teff$ $ and $\Delta n_{\rm x1}=(\numax - \numin_1)/\braket{\Dnu}$.}
Now we compute \numax$ $ { of the {\small MESA}  models} not from equation ({1}) but from the following new relation 
\begin{equation}
\frac{\nu_{\rm max}}{\nu_{\rm max\odot}}= f_\nu \left(\frac{T_{\rm eff\odot}}{T_{\rm eff}}\frac{\Gamma_{\rm \negthinspace 1s}}{\Gamma_{\rm \negthinspace 1s\odot}}\right)^{1/2} \frac{g}{g_{\odot}}.
\end{equation}
For the {\small MESA} models, the relation between \teff$ $ and $\Delta n_{\rm x1}$ is plotted in Fig. 11. 
From this relation, we obtain 
\begin{equation}
\frac{T_{\rm sis1}(\Delta n_{\rm x1})}{\rm T_{\rm eff \sun}}=1.157-1.093\times 10^{-9}(\Delta n_{\rm x1}+15)^{6.4}.
\end{equation}
The maximum difference between $T_{\rm sis1}$ from {equation ({15})} and model \teff$ $
is {generally} less than 100 K.
\begin{figure}
\includegraphics[width=101mm,angle=0]{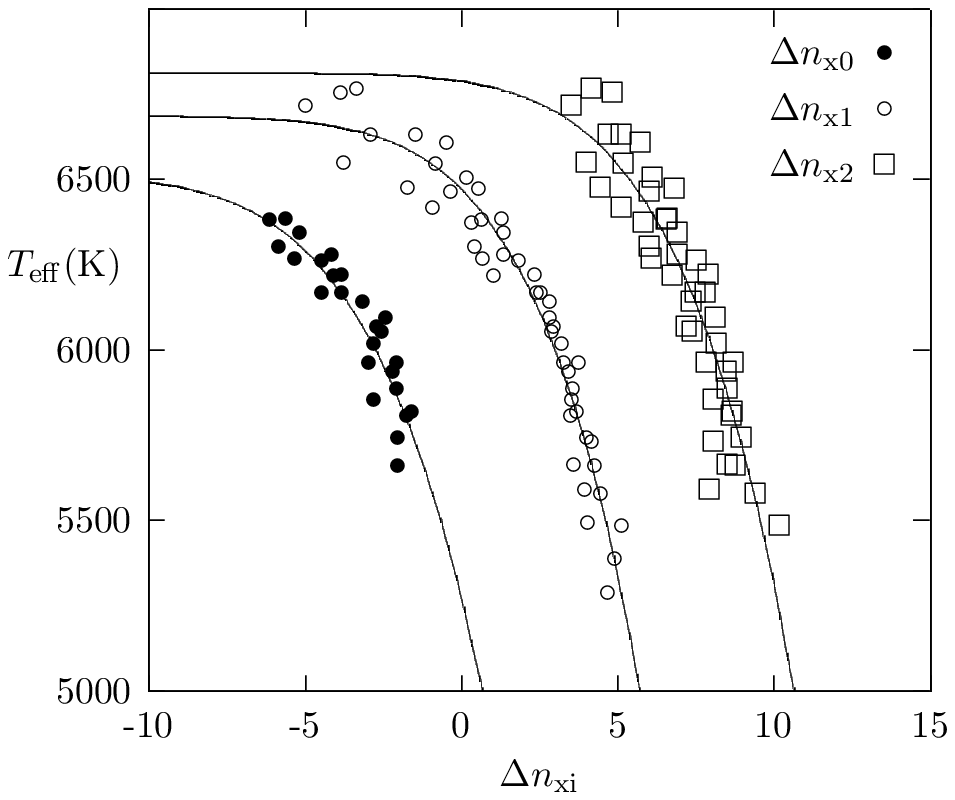}
\caption{Effective temperature is plotted \wrt $\Delta n_{\rm x0}$ (filled circles), $\Delta n_{\rm x1}$ (circles) and $\Delta n_{\rm x2}$ (squares). 
The fitting curves are given in equations { (15)-(17)}.
}
\end{figure}

Similarly, we derive fitting formula for effective temperatures in terms of min0 ($T_{\rm sis0}$) and min2 ($T_{\rm sis2}$) frequencies (see Fig. 11), more precisely, 
$\Delta n_{\rm x0}=(\numax - \numin_0)/\braket{\Dnu}$  and $\Delta n_{\rm x2}=(\numax - \numin_2)/\braket{\Dnu}$; 
\begin{equation}
\frac{T_{\rm sis0}(\Delta n_{\rm x0})}{\rm T_{\rm eff \sun}}=1.127-3.339\times 10^{-9}(\Delta n_{\rm x0}+20)^{6}
\end{equation}
and 
\begin{equation}
\frac{T_{\rm sis2}(\Delta n_{\rm x2})}{\rm T_{\rm eff \sun}}=1.179-4.049\times 10^{-9}(\Delta n_{\rm x2}+10)^{6}.
\end{equation}

Expression for typical uncertainty in $T_{\rm sis0}$ can be obtained from equation ({16}) as
\begin{equation}
\frac{\Delta T_{\rm sis0}}{T_{\rm eff\odot}}=2.0\times 10^{-8}(\Delta n_{\rm x0}+20)^{5}
\frac{\Delta \numax+\Delta \numin_0}{\braket\Dnu}.
\end{equation}
In a similar manner, expressions for uncertainties in $T_{\rm sis1}$ and $T_{\rm sis2}$ are derived from  { equations (15) and (17)}, respectively.  

\section{Results and Discussions}
\subsection{Effective temperatures of the target stars}
In the previous section, 
{effective temperature was shown to be} a function of order difference between 
the minima in ${\Dnu}$-$\nu$ graph and \numax. 
{Computations of $T_{\rm sis0}$, $T_{\rm sis1}$ and $T_{\rm sis2}$ of  the 
target stars} using { equations (15)-(17)} are listed in Table A1.  
$T_{\rm sis0}$, $T_{\rm sis1}$  and $T_{\rm sis2}$  are plotted \wrt$ $ $T_{\rm eS}$ in Fig. 12. 
In general, there is a very {close} agreement between asteroseismic \teff s ($T_{\rm sis0}$, $T_{\rm sis1}$  and $T_{\rm sis2}$)  and $T_{\rm eS}$. 
However, {a discrepancy occurs} between $T_{\rm eS}$ and $T_{\rm sis1}$ {in the case where} $T_{\rm eS}< 5500$ K. 
min2 is only seen in $\braket{\Dnu}$-$\nu$ graph of very hot solar-like oscillating stars. { Therefore, } 
{it was only possible to find $T_{\rm sis2}$ for 15 stars.}

For some stars, there is a systematic difference between asteroseismic and non-asteroseismic \teff s. 
{In some cases, such as  KIC 3427720, KIC 3544595 and KIC 5866724, $T_{\rm eS}$ is greater than both $T_{\rm sis0}$ and $T_{\rm sis1}$,
therefore, the the values of \numax$ $ should be decreased to fit $T_{\rm sis0}$ to $T_{\rm eS}$.}
For some targets (for example KIC 3424541, KIC 6679371 and KIC 7799349),
however,
$T_{\rm eS}$ is less than both of $T_{\rm sis0}$ and $T_{\rm sis1}$. In this case, \numax$ $ might be increased to fit $T_{\rm sis0}$ to $T_{\rm eS}$.
This method will be considered in the next paper of this series.
%
\begin{figure}
\includegraphics[width=101mm,angle=0]{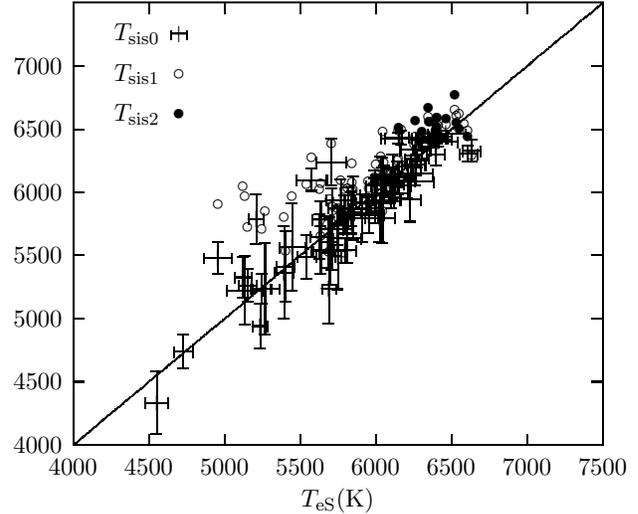}
\caption{Effective temperatures { of the target stars} obtained by using oscillation frequencies are plotted \wrt $T_{\rm eS}$.
}
\end{figure}

For the Sun, the results are very {impressive; the mean effective temperature} from min0, min1 and min2 {was} 
obtained as $5804 \pm 60$ K. This value is very close to observed effective temperature of the Sun ($5777$ K), {with very low standard deviation.}
Since there is no calibration of asteroseismic relations for effective temperature ({equations 15-17}), {this finding alone shows
the great value of asteroseismic tools.}

\subsection{Masses and radii}
\begin{figure}
\includegraphics[width=101mm,angle=0]{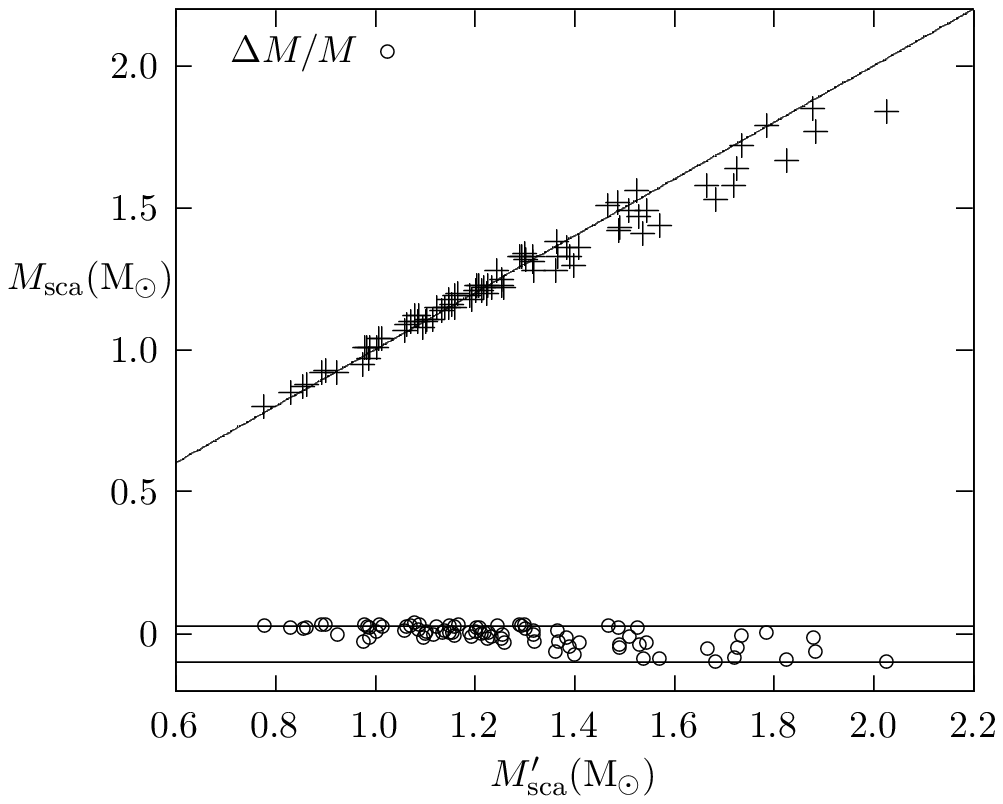}
\caption{Mass of target stars computed by using the tuned scaling relation  (equation { 9}) is plotted \wrt mass from the conventional scaling relation.
}
\end{figure}
In Fig. 13, 
{$M_{\rm sca}$ ((equation  9) ) is plotted with respect to $M'_{\rm sca}$.} 
Also shown in Fig. 13 is the fractional difference between $M_{\rm sca}$ and $M'_{\rm sca}$. The horizontal solid lines are for 0.03 and -0.10.
For  $M < 1.3$ \MSbit, $M_{\rm sca}$ is nearly 3 per cent greater than $M'_{\rm sca}$. For $M > 1.3$ \MSbit; {however, the
difference between $M_{\rm sca}$ and $M'_{\rm sca}$ does not rise above 10 per cent.}

{ $M_{\rm sca}$, $M'_{\rm sca}$, $R_{\rm sca}$ and $R'_{\rm sca}$ are listed in Table A1, together with the } masses ($M_{\rm lit}$) and radii 
($R_{\rm lit}$) of the target stars compiled from the literature.
{ $M_{\rm lit}$ and $R_{\rm lit}$} {were} found by constructing interior models of the target stars,
except { Procyon A}. The mass and radius of { Procyon A} given in Table A1 {were} obtained from observations (see Section 4.3).
In Fig. 14, fractional mass difference between mass  found from new scaling relation and mass from the literature,  
${\Delta M}/{M}=(M_{\rm lit}-M_{\rm sca})/M_{\rm sca}$, is plotted \wrt
fractional radius difference, ${\Delta R}/{R}=(R_{\rm lit}-R_{\rm sca})/R_{\rm sca}$. 
This figure shows that the maximum difference { between $R_{\rm sca}$ and $R_{\rm lit}$} is about 10 {per cent, and} is about 25 per cent for the mass.
{More crucially, it can be seen that  there is a linear relation between the fractional mass and radius:
${\Delta M}/{M}=3 {\Delta R}/{R}$, imlpying that} the density of the models  are the same as that of given  by $M_{\rm sca}/R_{\rm sca}^3$.
\begin{figure}
\includegraphics[width=101mm,angle=0]{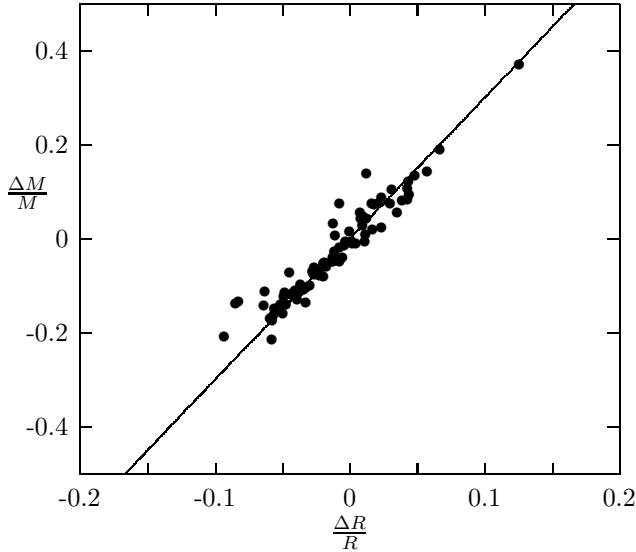}
\caption{Fractional mass difference is plotted \wrt fractional radius difference.
The solid line represents $\Delta M/M = 3 \Delta R/R$. 
}
\end{figure}
\begin{figure}
\includegraphics[width=101mm,angle=0]{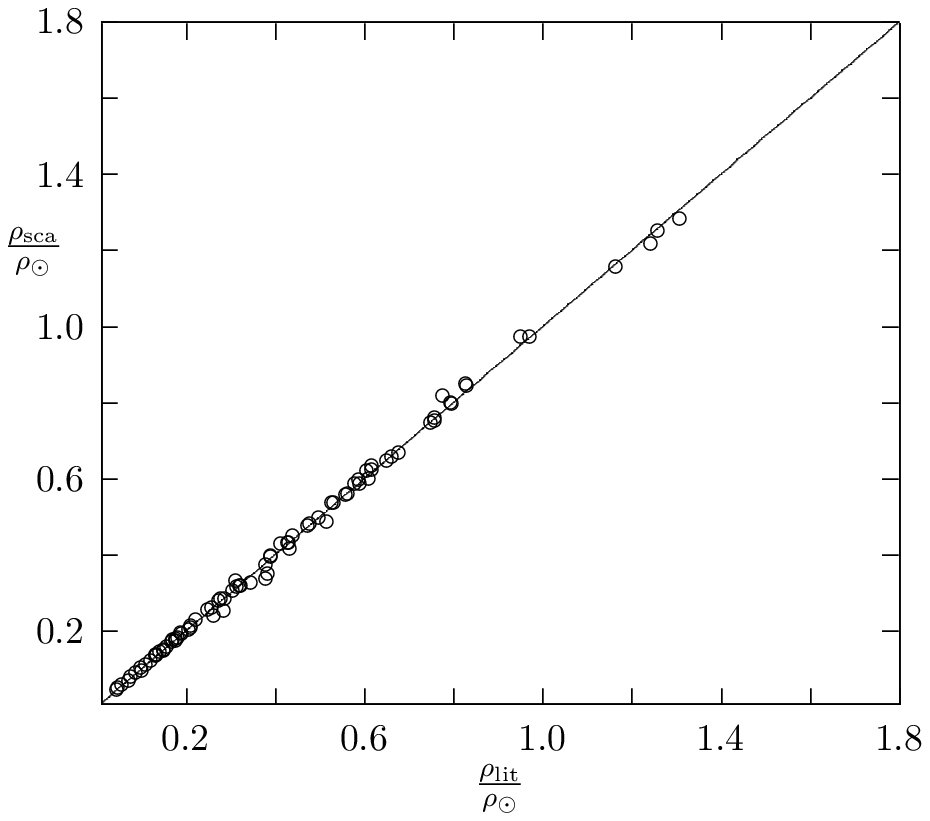}
\caption{$\rho_{\rm sca}$ is plotted \wrt $\rho_{\rm lit}$.
}
\end{figure}
This result arises from the fact that oscillation frequency of a given mode depends on mean stellar {density, and may imply that
density is indeed the fitted parameter in the calibration process of model frequencies to the observed oscillation frequencies.}
In that case, 
{it is possible that different combinations of $M$ and 
$R$ yield the same density.} In Fig. 15, $\rho_{\rm sca}/\rho_{\sun}=M_{\rm sca}/R^3_{\rm sca}$ is plotted \wrt 
$\rho_{\rm lit}/\rho_{\sun}=M_{\rm lit}/R^3_{\rm lit}$ (masses and radii are in solar units).
Despite the significant differences between the masses and the radii (Fig. 14), the densities $\rho_{\rm sca}$ 
and $\rho_{\rm lit}$ are in very good agreement.
Therefore, 
{our further research will address the need for further constraints, such as frequencies of minimum $\Dnu$, when
applying asteroseismic methods for finding fundamental
properties of stars.}

\subsection{Comparison with masses and radii from non-asteroseismic measurements}
Asteroseismic and non-asteroseismic effective temperatures of the Sun are compared above. 
Among the stars 
{focused on,  Procyon A is only star for which mass and radius are found by non-asteroseismic methods.} 
Its mass {was} obtained by Bond et al. (2015) as { 1.478 $\pm$ 0.012 \MSbit}, and its radius {was} found as { 2.03 $\pm$ 0.013 \RS}
(Aufdenberg, Ludwig \& Kervella,  2005). 
{Conventional scaling relations give mass and radius as 1.63 \MS and 2.14 \RSbit, respectively.
Results from the new scaling relations for these two quantities are 1.46 \MS and 2.03 \RSbit, respectively, 
very close to the observed values, showing the
effectiveness of these new relations.}

Another non-asteroseismic observational result for testing asteroseismic methods of the present study {pertains} to HD 181907.
Its {radius, found by interferometric methods, is $12.1\pm0.5$ \RS (Baines et al. 2014), and 
its} literature value is $12.93\pm0.95$ \RS (Ghezzi \& Johnson 2015). The old and new scaling relations, however, 
yield $12.91$ and  $12.42$ \RSbit, respectively. Again, the radius from the new scaling relation
is in very good agreement with the interferometric radius.

\section{Conclusions}
{ Asteroseismology of solar-like oscillating stars
provides information about fundamental properties of these stars} through the scaling relations. In these relations it is assumed that 
the first adiabatic exponent $\Gamma_{\rm \negthinspace 1s}$ is constant. However, analysis of our models constructed by the {\small MESA} 
code shows that $\Gamma_{\rm \negthinspace 1s}$ significantly changes through the surfaces of solar-like oscillating stars, depending on effective temperature.
Furthermore, the ratio of the { mean} large separation $\braket{\Dnu}$ to $\braket{\rho}^{0.5}$, which is customarily assumed to be constant,
is a linear function of $\Gamma_{\rm \negthinspace 1s}$. Thus, it seems that $\Gamma_{\rm \negthinspace 1s}$ is the 
{factor in scaling relation that has previously been overlooked.} 

In contrast to the literature, we do not consider the ratios of $\nu_{\rm ac}/\numax $  and $\nu_{\rm ac}/(g/T_{\rm eff}^{0.5}) $ as { constant}. We show that 
these ratios can also be taken as functions of $\Gamma_{\rm \negthinspace 1s}$. Then, 
we revise the scaling relations and obtain new relations for stellar mass and radius { (equations 9 and 10)}, and then compute mass and radius of 
{\it Kepler} and {\it CoRoT} targets (89 stars + the Sun) using their asteroseismic properties. 
{A difference of up to 10 per cent exists between the masses from new and old scaling relations.}
However, the great part of the uncertainties in mass and radius found from scaling relations comes from
uncertainty in $\numax$. 

When {mass and radius obtained from scaling relations are compared with those available in the literature, 
significant differences are seen}. The fractional differences between the masses are up to $\Delta M/M = 0.25$ and $\Delta R/R = 0.10$.
{Particularly noteworthy, however, is the  linear relation between $\Delta M/M$ and $\Delta R/R$: $\Delta M/M = 3 \Delta R/R$ { (see Fig. 14)},
highlighting that the fitted parameter is indeed  mean density when oscillation frequencies of interior models are fitted to  the observed oscillation frequencies.}

We also {computed effective temperatures of these stars using} purely asteroseismic methods. In our previous paper (Paper I), 
{the effective temperature of models was shown to be a function of $\Delta n_{\rm x1}$, which is approximately}
the order difference between the { frequencies} of maximum amplitude and min1 in the $\Dnu - \nu$ graph. 
{Taking} into account the effect of $\Gamma_{\rm \negthinspace 1s}$, we derive new relations between $T_{\rm eff}$ and $\Delta n_{\rm x1}$ by using 
the {\small MESA} models. Similar expressions are obtained for $\Delta n_{\rm x0}$ and $\Delta n_{\rm x2}$. 
{This allows us in principle to use three different methods to compute $T_{\rm eff}$} in terms of the  oscillation 
frequencies of the target stars.
These effective temperatures ($T_{\rm sis0}$, $T_{\rm sis1}$  and $T_{\rm sis2}$; see { equations 15-17})
are in general in very good agreement within themselves and with $T_{\rm eff}$ from conventional methods. {A significant} difference 
appears between $T_{\rm sis1}$ and $T_{\rm eS}$ {for cool stars, but not for the Sun, for example.} 
{The value of these new approaches lies in the increased number of methods they allow for computing the effective temperature of a solar-like oscillating star.
The six methods consist of three asteroseismic, one spectroscopic and two
photometric methods.}

In principle, we can compute the fundamental stellar parameters by purely asterosesimic quantities,
provided that the oscillation frequencies are precisely determined. The solar effective temperature is found 
as 5742, 5831 and 5840 K from frequencies of min0, min1 and min2, respectively. All of these values are very close to 
$T_{\rm eff \odot}$. Another {key} result we obtain is about { Procyon A}. Its mass is determined by using asterometric data 
from $Hubble$ $Space$ $Telescope$ as $1.478\pm0.012$ \MSbit. While the conventional scaling relation yields 1.63 \MSbit, the new scaling relation
gives {a much more accurate figure,} 1.46 \MSbit.

However, in some cases, we confirm that there are systematic differences between asteroseismic and non-asteroseismic effective temperatures.
These differences can be {reduced or eliminated} by increasing or decreasing \numax. Such a 
{modification, ciritical} for obtaining more precise mass and radius from the scaling {relations,  will} be the subject of our 
next paper.


\section*{Acknowledgements}
Dr. Eric Michel is acknowledged for his discussions.
This work is supported by the Scientific and Technological Research Council of Turkey (T\"UB\.ITAK: 112T989).

\appendix
\onecolumn
\section{Basic asteroseismic and non-asteroseismic properties of {\it Kepler} and {\it CoRoT} Targets}
\small\addtolength{\tabcolsep}{-2pt}

\begin{landscape}

\begin{longtable}{@{}rrrrrrrrrrrrrrrrrrrrrrr@{}}
\caption[Continued.]{ 
Basic properties of target stars. Columns are organized as star name, frequency of maximum amplitude, reference frequencies for min0, min1 and min2, mean large and small separations between 
oscillation frequencies, effective temperatures (from spectra, { $V-K$ and $B-V$} colours (see Section 2) and from minima min0, min1 and min2 (see Section 3), respectively), masses and radii 
(from new and conventional scaling relations (see Section 3) and from literature, respectively), surface gravities (from new and conventional scaling relations (see Section 3), 
and from spectra, respectively), 
and numbers of references. Second row describes uncertainties of these basic properties. Unlike others, for { Procyon A}, mass and radius are the observed values, not the model values in the 
literature. Since $T_{\rm eS}$, $T_{\rm eVK}$ and $T_{\rm eBV}$  of KIC 11771760 are not available, we use $T_{\rm sis0}$ in scaling relations. 
The Sun is given 
at the end of the table. 
}\\
\hline
Star    & \numax & $\nu_{\rm min0}$ & $\nu_{\rm min1}$ & $\nu_{\rm min2}$ &  $\braket{\Dnu}$ & $\braket{\delta \nu_{02}}$ & $T_{\rm eS}$ &  $T_{\rm eVK}$ &  $T_{\rm eBV}$ & $T_{\rm sis0}$ & $T_{\rm sis1}$ & $T_{\rm sis2}$ & $M_{\rm sca}$ &$ M'_{\rm sca}$ & $M_{\rm lit}$ & $R_{\rm sca}$ &$ R'_{\rm sca}$ & $R_{\rm lit}$  & $\log g_{\rm sca}$ &$ \log g'_{\rm sca}$ & $\log g_{\rm spc}$ &  Ref \\
		& \muHz & \muHz   & \muHz  & \muHz  &  \muHz & \muHz & K &  K   &  K   & K   &  K   & K   & \MS     &  \MS     &  \MS   & \RS     &  \RS   &  \RS   &  & &  &   \\
\hline
\endfirsthead
\caption[-- continued from previous page]{-- continued from previous page}\\
\hline
Star    & \numax & $\nu_{\rm min0}$ & $\nu_{\rm min1}$ & $\nu_{\rm min2}$ &  $\braket{\Dnu}$ & $\braket{\delta \nu_{02}}$ & $T_{\rm eS}$ &  $T_{\rm eVK}$ &  $T_{\rm eBV}$ & $T_{\rm sis0}$ & $T_{\rm sis1}$ & $T_{\rm sis2}$ & $M_{\rm sca}$ &$ M'_{\rm sca}$ & $M_{\rm lit}$ & $R_{\rm sca}$ &$ R'_{\rm sca}$ & $R_{\rm lit}$  & $\log g_{\rm sca}$ &$ \log g'_{\rm sca}$ & $\log g_{\rm spc}$ &  Ref \\
		& \muHz & \muHz   & \muHz  & \muHz  &  \muHz & \muHz & K &  K   &  K   & K   &  K   & K   & \MS     &  \MS     &  \MS   & \RS     &  \RS   &  \RS   &  & &  &   \\

\hline
\endhead

\hline
\endfoot

\bf{ 1435467} &   1324.0 &   1626.4 &   1274.0 &      --- &   70.9 &    4.8 &   6264 &   6224 &   6587 &   6217 &   6398 &      --- &     1.18 &     1.25 &     1.27 &     1.62 &     1.66 &     1.64 &     4.09 &     4.09 &     4.09 & 2,13,17 \\
              &     39.7 &     16.3 &     12.7 &      --- &    0.8 &    --- &     60 &     51 &    185 &     82 &     87 &      --- &     0.18 &     0.18 &     0.05 &     0.09 &     0.09 &     0.03 &     0.01 &     0.01 &     0.03 &   39    \\
\bf{ 2837475} &   1630.0 &   2265.6 &   1689.9 &   1276.8 &   75.2 &    6.7 &   6462 &   6545 &   6488 &   6464 &   6534 &   6575 &     1.75 &     1.93 &     1.39 &     1.76 &     1.85 &     1.59 &     4.19 &     4.19 &     3.95 & 2,17,39 \\
              &     54.0 &     22.7 &     16.9 &     12.8 &    1.3 &    --- &    125 &     54 &    144 &     22 &     72 &    101 &     0.35 &     0.35 &     0.06 &     0.14 &     0.14 &     0.03 &     0.02 &     0.02 &     0.23 &   40    \\
\bf{ 3424541} &    745.0 &   1046.6 &    755.4 &      --- &   41.1 &    4.7 &   6165 &   6249 &   6322 &   6431 &   6493 &      --- &     1.85 &     1.92 &     1.64 &     2.70 &     2.76 &     2.53 &     3.84 &     3.84 &     3.90 & 2,17,39 \\
              &     55.0 &     10.5 &      7.6 &      --- &    1.1 &    --- &    108 &     73 &    249 &     56 &    137 &      --- &     0.65 &     0.65 &     0.04 &     0.37 &     0.37 &     0.07 &     0.02 &     0.02 &     0.21 &   40    \\
\bf{ 3427720} &   2756.0 &   3044.3 &   2325.2 &      --- &  120.0 &   10.3 &   6040 &   6038 &   6055 &   5937 &   5845 &      --- &     1.27 &     1.30 &     1.13 &     1.17 &     1.19 &     1.13 &     4.40 &     4.40 &     4.38 & 2,13,17 \\
              &    191.0 &     30.4 &     23.3 &      --- &    2.0 &    --- &     60 &     60 &    211 &    338 &    436 &      --- &     0.37 &     0.37 &     0.04 &     0.13 &     0.13 &     0.01 &     0.02 &     0.02 &     0.03 &   39    \\
\bf{ 3544595} &   3366.0 &   3350.9 &   2702.9 &      --- &  145.5 &    8.7 &   5689 &   5640 &   5434 &   5237 &   5523 &      --- &     1.01 &     1.00 &     0.91 &     0.96 &     0.96 &     0.92 &     4.48 &     4.48 &     4.56 & 4,29,46 \\
              &     81.0 &     33.5 &     27.0 &      --- &    1.5 &    --- &     48 &     52 &    200 &    280 &    225 &      --- &     0.13 &     0.13 &     0.06 &     0.05 &     0.05 &     0.02 &     0.01 &     0.01 &     0.06 &         \\
\bf{ 3632418} &   1159.0 &   1370.9 &   1055.0 &      --- &   60.4 &    3.8 &   6148 &   6154 &   6148 &   6122 &   6258 &      --- &     1.49 &     1.55 &     1.27 &     1.95 &     1.99 &     1.83 &     4.03 &     4.03 &     3.94 & 2,17,39 \\
              &     44.0 &     13.7 &     10.6 &      --- &    0.4 &    --- &    111 &     38 &     77 &    126 &    139 &      --- &     0.25 &     0.25 &     0.03 &     0.12 &     0.12 &     0.03 &     0.01 &     0.01 &     0.21 &   40    \\
\bf{ 3656476} &   1887.0 &      --- &      --- &      --- &   93.2 &    4.4 &   5710 &   5752 &   5303 &      --- &      --- &      --- &     1.06 &     1.05 &     1.09 &     1.31 &     1.31 &     1.32 &     4.23 &     4.23 &     4.23 & 13,17,37\\
              &     40.0 &      --- &      --- &      --- &    1.3 &    --- &     60 &     52 &    137 &      --- &      --- &      --- &     0.14 &     0.14 &     0.01 &     0.07 &     0.07 &     0.03 &     0.01 &     0.01 &     0.03 &         \\
\bf{ 3733735} &   1974.0 &      --- &   2211.0 &   1480.0 &   91.6 &    9.9 &   6548 &   6610 &   6581 &      --- &   6620 &   6499 &     1.42 &     1.59 &     1.32 &     1.43 &     1.52 &     1.37 &     4.28 &     4.28 &     3.99 & 2,17,39 \\
              &    121.0 &      --- &     22.1 &     14.8 &    2.5 &    --- &    156 &     43 &    151 &      --- &     68 &    204 &     0.47 &     0.47 &     0.04 &     0.18 &     0.18 &     0.02 &     0.02 &     0.02 &     0.22 &   40    \\
\bf{ 3735871} &   2633.0 &   2850.7 &      --- &      --- &  124.7 &   12.3 &   5908 &   6207 &   5908 &   5797 &      --- &      --- &     0.93 &     0.94 &     1.07 &     1.03 &     1.04 &     1.09 &     4.38 &     4.38 &     --- & 2,17    \\
              &     79.0 &     28.5 &      --- &      --- &    3.3 &    --- &    100 &     49 &    202 &    189 &      --- &      --- &     0.21 &     0.21 &     0.12 &     0.09 &     0.09 &     0.05 &     0.02 &     0.02 &     --- &         \\
\bf{ 4349452} &   2106.0 &   2568.7 &   1884.4 &      --- &   97.6 &    7.7 &   6270 &   6194 &   6048 &   6267 &   6160 &      --- &     1.33 &     1.40 &     1.19 &     1.36 &     1.40 &     1.31 &     4.30 &     4.29 &     4.28 & 9,29,36 \\
              &     50.0 &     25.7 &     18.8 &      --- &    1.0 &    --- &     79 &    107 &      --- &     69 &    125 &      --- &     0.17 &     0.17 &     0.06 &     0.07 &     0.07 &     0.02 &     0.01 &     0.01 &     0.03 &         \\
\bf{ 4914923} &   1849.0 &   1947.8 &      --- &      --- &   88.7 &    6.1 &   5808 &   5721 &   5910 &   5635 &      --- &      --- &     1.24 &     1.24 &     1.10 &     1.43 &     1.43 &     1.37 &     4.22 &     4.22 &     4.28 & 17,37,40\\
              &     46.0 &     19.5 &      --- &      --- &    0.3 &    --- &     92 &     65 &    194 &    192 &      --- &      --- &     0.14 &     0.14 &     0.01 &     0.06 &     0.06 &     0.05 &     0.01 &     0.01 &     0.21 &         \\
\bf{ 5184732} &   2068.0 &   2182.6 &   1705.5 &      --- &   95.1 &    5.9 &   5840 &   5836 &   5611 &   5660 &   5779 &    --- &     1.32 &     1.32 &     1.25 &     1.39 &     1.39 &     1.36 &     4.27 &     4.27 &     4.26 & 13,17,37\\
              &     47.0 &      --- &     18.8 &      --- &    1.3 &    --- &     60 &     41 &     91 &      --- &    113 &      --- &     0.18 &     0.18 &     0.01 &     0.08 &     0.08 &     0.01 &     0.01 &     0.01 &     0.03 &         \\
\bf{ 5512589} &   1224.0 &      --- &      --- &      --- &   68.2 &    5.5 &   5764 &   5687 &   5583 &      --- &      --- &      --- &     1.02 &     1.02 &     1.16 &     1.60 &     1.59 &     1.67 &     4.04 &     4.04 &     4.22 & 17,37,40\\
              &     43.0 &      --- &      --- &      --- &    0.7 &    --- &     95 &     64 &    203 &      --- &      --- &      --- &     0.17 &     0.17 &     0.01 &     0.10 &     0.10 &     0.01 &     0.01 &     0.01 &     0.21 &         \\
\bf{ 5607242} &    610.0 &    745.3 &    543.4 &      --- &   40.5 &    3.7 &   5572 &   5572 &   5070 &   6097 &   6270 &      --- &     0.97 &     0.96 &     1.33 &     2.22 &     2.21 &     2.49 &     3.73 &     3.73 &     --- & 2,17    \\
              &     18.3 &      7.5 &      5.4 &      --- &    0.8 &    --- &    100 &     83 &    272 &     88 &     88 &      --- &     0.19 &     0.19 &     0.11 &     0.17 &     0.17 &     0.10 &     0.02 &     0.02 &     --- &         \\
\bf{ 5689820} &    695.0 &      --- &      --- &      --- &   41.0 &    3.9 &   4978 &      --- &      --- &      --- &      --- &      --- &     1.11 &     1.15 &     1.14 &     2.29 &     2.33 &      --- &     3.76 &     3.76 &     --- & 22      \\
              &     15.0 &      --- &      --- &      --- &    0.5 &    --- &    167 &      --- &      --- &      --- &      --- &      --- &     0.18 &     0.18 &      ---  &     0.14 &     0.14 &      --- &     0.01 &     0.01 &     --- &         \\
\bf{ 5866724} &   1880.0 &   2171.6 &   1674.4 &      --- &   89.6 &    6.6 &   6211 &   6410 &   5574 &   6085 &   6155 &      --- &     1.32 &     1.38 &     1.27 &     1.44 &     1.47 &     1.42 &     4.24 &     4.24 &     4.23 & 16,29   \\
              &     60.0 &     21.7 &     16.7 &      --- &    0.9 &    --- &    167 &    104 &    369 &    130 &    153 &      --- &     0.23 &     0.23 &     0.06 &     0.09 &     0.09 &     0.02 &     0.01 &     0.01 &     0.01 &         \\
\bf{ 5955122} &    861.0 &    952.7 &    717.9 &      --- &   49.4 &    4.8 &   5952 &   5917 &   6068 &   5822 &   6026 &      --- &     1.33 &     1.35 &     1.12 &     2.16 &     2.17 &     2.04 &     3.89 &     3.89 &     4.13 & 2,17,39 \\
              &     24.0 &      9.5 &      7.2 &      --- &    0.9 &    --- &    100 &     78 &    192 &    144 &    130 &      --- &     0.24 &     0.24 &     0.05 &     0.16 &     0.16 &     0.03 &     0.02 &     0.02 &     0.21 &   40    \\
\bf{ 6106415} &   2260.0 &   2533.1 &   1909.2 &      --- &  103.9 &    6.5 &   5990 &   6056 &   6040 &   5980 &   5904 &      --- &     1.24 &     1.26 &     1.12 &     1.28 &     1.29 &     1.24 &     4.32 &     4.32 &     4.31 & 13,17,37\\
              &     53.0 &     25.3 &     19.1 &      --- &    0.3 &    --- &     60 &     50 &     76 &    129 &    161 &      --- &     0.12 &     0.12 &     0.02 &     0.04 &     0.04 &     0.01 &     0.01 &     0.01 &     0.03 &         \\
\bf{ 6116048} &   2020.0 &   2250.4 &   1748.1 &      --- &  100.5 &    5.9 &   5991 &   6109 &   5844 &   5916 &   6069 &      --- &     1.01 &     1.03 &     1.12 &     1.23 &     1.24 &     1.26 &     4.27 &     4.27 &     4.09 & 2,3,37  \\
              &     60.6 &     22.5 &     17.5 &      --- &    0.2 &    --- &    124 &     41 &    119 &    156 &    154 &      --- &     0.13 &     0.13 &     0.02 &     0.06 &     0.06 &     0.01 &     0.01 &     0.01 &     0.22 &   40    \\
\bf{ 6508366} &    926.0 &   1267.5 &    978.3 &    672.2 &   51.5 &    3.3 &   6354 &   6268 &   6414 &   6400 &   6548 &   6551 &     1.46 &     1.57 &     1.36 &     2.14 &     2.22 &     2.08 &     3.94 &     3.94 &     3.94 & 2,13,17 \\
              &     36.0 &     12.7 &      9.8 &      6.7 &    0.8 &    --- &     60 &     60 &    177 &     43 &     64 &    101 &     0.28 &     0.28 &     0.04 &     0.16 &     0.16 &     0.02 &     0.01 &     0.01 &     0.03 &   39    \\
\bf{ 6603624} &   2402.0 &   2529.7 &   2080.5 &      --- &  109.7 &    5.5 &   5625 &   5642 &   5364 &   5649 &   6017 &      --- &     1.12 &     1.11 &     1.09 &     1.20 &     1.19 &     1.18 &     4.33 &     4.33 &     4.32 & 2,13,17 \\
              &     51.0 &     25.3 &     20.8 &      --- &    1.7 &    --- &     60 &     58 &    113 &    179 &    137 &      --- &     0.16 &     0.16 &     0.03 &     0.07 &     0.07 &     0.02 &     0.01 &     0.01 &     0.03 &   39    \\
\bf{ 6679371} &    908.0 &   1284.8 &   1000.6 &    725.8 &   50.6 &    4.1 &   6344 &   6375 &   6453 &   6435 &   6591 &   6662 &     1.48 &     1.59 &     1.56 &     2.17 &     2.25 &     2.19 &     3.93 &     3.93 &     3.92 & 2,3,39  \\
              &     27.2 &     12.8 &     10.0 &      7.3 &    0.7 &    --- &    131 &     21 &    132 &     26 &     41 &     57 &     0.26 &     0.26 &     0.03 &     0.15 &     0.15 &     0.02 &     0.01 &     0.01 &     0.21 &   40    \\
\bf{ 6933899} &   1391.0 &   1538.7 &   1178.0 &      --- &   71.8 &    4.9 &   5837 &   5700 &   5982 &   5866 &   6009 &      --- &     1.24 &     1.24 &     1.14 &     1.64 &     1.64 &     1.60 &     4.10 &     4.10 &     4.21 & 2,17,39 \\
              &     32.0 &     15.4 &     11.8 &      --- &    1.0 &    --- &     97 &     58 &    191 &    133 &    128 &      --- &     0.18 &     0.18 &     0.03 &     0.10 &     0.10 &     0.02 &     0.01 &     0.01 &     0.22 &   40    \\
\bf{ 7103006} &   1124.0 &   1432.9 &   1134.4 &    790.3 &   60.1 &    4.5 &   6394 &   6351 &   6151 &   6302 &   6486 &   6480 &     1.41 &     1.53 &     1.43 &     1.90 &     1.98 &     1.90 &     4.03 &     4.03 &     4.01 & 2,13,17 \\
              &     54.0 &     14.3 &     11.3 &      7.9 &    1.1 &    --- &     60 &     42 &    126 &     89 &    100 &    147 &     0.33 &     0.33 &     0.05 &     0.17 &     0.17 &     0.03 &     0.02 &     0.02 &     0.03 &   39    \\
\bf{ 7106245} &   2323.0 &      --- &   2105.0 &      --- &  111.6 &    7.0 &   6000 &   6000 &   5725 &      --- &   6218 &      --- &     1.01 &     1.03 &      ---  &     1.14 &     1.15 &      --- &     4.33 &     4.33 &     --- & 2       \\
              &     69.7 &      --- &     21.0 &      --- &    1.1 &    --- &     99 &     98 &    479 &      --- &    133 &      --- &     0.16 &     0.16 &      ---  &     0.07 &     0.07 &      --- &     0.01 &     0.01 &     --- &         \\
\bf{ 7206837} &   1592.0 &   2023.7 &   1545.1 &   1153.7 &   78.7 &    6.2 &   6304 &   6190 &   6480 &   6330 &   6411 &   6478 &     1.36 &     1.44 &     1.46 &     1.58 &     1.63 &     1.56 &     4.17 &     4.17 &     4.17 & 2,13,17 \\
              &     70.0 &     20.2 &     15.5 &     11.5 &    1.4 &    --- &     60 &     60 &    263 &     80 &    124 &    149 &     0.29 &     0.29 &     0.05 &     0.13 &     0.13 &     0.02 &     0.02 &     0.02 &     0.03 &   39    \\
\bf{ 7341231} &    408.0 &    384.5 &      --- &      --- &   28.8 &    3.4 &   5233 &   5440 &   5438 &   4941 &      --- &      --- &     1.03 &     1.03 &     0.90 &     2.84 &     2.84 &     2.69 &     3.54 &     3.54 &     3.54 & 2,17,21 \\
              &      8.0 &      3.8 &      --- &      --- &    0.7 &    --- &     50 &     51 &    180 &    174 &      --- &      --- &     0.17 &     0.17 &     0.10 &     0.21 &     0.21 &     0.20 &     0.02 &     0.02 &     0.03 &         \\
\bf{ 7680114} &   1684.0 &      --- &      --- &      --- &   85.1 &    --- &   5799 &   5891 &   5588 &      --- &      --- &      --- &     1.10 &     1.10 &     1.19 &     1.41 &     1.41 &     1.45 &     4.18 &     4.18 &     4.25 & 17,37,40\\
              &     47.0 &      --- &      --- &      --- &    1.3 &    --- &     91 &     94 &    268 &      --- &      --- &      --- &     0.19 &     0.19 &     0.01 &     0.09 &     0.09 &     0.03 &     0.01 &     0.01 &     0.21 &         \\
\bf{ 7747078} &    936.0 &   1039.3 &    792.5 &      --- &   53.4 &    4.7 &   5840 &   5754 &   5727 &   5840 &   6074 &      --- &     1.23 &     1.24 &     1.06 &     2.00 &     2.00 &     1.89 &     3.93 &     3.93 &     3.91 & 2,13,17 \\
              &     32.0 &     10.4 &      7.9 &      --- &    0.3 &    --- &     60 &     66 &    172 &    165 &    147 &      --- &     0.17 &     0.17 &     0.05 &     0.10 &     0.10 &     0.02 &     0.01 &     0.01 &     0.03 &   39    \\
\bf{ 7799349} &    561.0 &    580.6 &    448.6 &      --- &   33.2 &    3.4 &   4954 &   4962 &   4821 &   5479 &   5902 &      --- &     1.34 &     1.39 &     1.39 &     2.80 &     2.86 &      --- &     3.67 &     3.67 &     3.33 & 2,22,40 \\
              &      8.0 &      5.8 &      4.5 &      --- &    0.4 &    --- &     92 &     41 &    124 &    124 &     87 &      --- &     0.16 &     0.16 &      ---  &     0.13 &     0.13 &      --- &     0.01 &     0.01 &     0.22 &         \\
\bf{ 7871531} &   3344.0 &   3403.0 &   2658.3 &      --- &  151.3 &   10.1 &   5400 &   5289 &   5641 &   5413 &   5533 &      --- &     0.78 &     0.78 &     0.84 &     0.86 &     0.86 &     0.87 &     4.46 &     4.46 &     4.49 & 2,13,39 \\
              &    100.3 &     34.0 &     26.6 &      --- &    3.6 &    --- &     60 &     51 &    145 &    279 &    253 &      --- &     0.16 &     0.16 &     0.02 &     0.07 &     0.07 &     0.01 &     0.02 &     0.02 &     0.20 &         \\
\bf{ 7976303} &    851.0 &   1036.3 &    754.0 &      --- &   51.0 &    4.5 &   6053 &   5967 &   6315 &   6139 &   6228 &      --- &     1.15 &     1.17 &     1.17 &     2.00 &     2.03 &     2.03 &     3.89 &     3.89 &     3.87 & 13,17,37\\
              &     20.0 &     10.4 &      7.5 &      --- &    0.6 &    --- &     60 &     66 &    182 &     75 &     87 &      --- &     0.15 &     0.15 &     0.02 &     0.10 &     0.10 &     0.05 &     0.01 &     0.01 &     0.03 &         \\
\bf{ 8006161} &   3481.0 &   3518.3 &   2922.7 &      --- &  149.2 &   10.3 &   5390 &   5378 &   5189 &   5365 &   5800 &      --- &     0.93 &     0.93 &     1.00 &     0.92 &     0.92 &     0.93 &     4.48 &     4.48 &     4.49 & 2,13,17 \\
              &    133.0 &     35.2 &     29.2 &      --- &    1.8 &    --- &     60 &      --- &      --- &    367 &    275 &      --- &     0.17 &     0.17 &     0.01 &     0.06 &     0.06 &     --- &     0.01 &     0.01 &     0.03 &   37    \\
\bf{ 8026226} &    545.0 &    687.7 &    479.1 &      --- &   34.6 &    3.6 &   6230 &   6233 &   6204 &   6202 &   6227 &      --- &     1.45 &     1.53 &     1.50 &     2.80 &     2.87 &     2.75 &     3.71 &     3.71 &     3.71 & 2,13,17 \\
              &     22.0 &      6.9 &      4.8 &      --- &    0.6 &    --- &     60 &     45 &    127 &     91 &    125 &      --- &     0.30 &     0.30 &     0.03 &     0.22 &     0.22 &     0.04 &     0.02 &     0.02 &     0.03 &   39    \\
\bf{ 8219268} &    109.0 &     90.5 &      --- &      --- &    9.4 &    1.1 &   4550 &   4424 &   4702 &   4335 &      --- &      --- &     1.27 &     1.42 &     1.34 &     6.32 &     6.69 &     6.53 &     2.94 &     2.94 &     3.00 & 29,33   \\
              &      3.3 &      0.9 &      --- &      --- &    0.1 &    --- &     75 &      --- &      --- &    248 &      --- &      --- &     0.20 &     0.20 &     0.17 &     0.37 &     0.37 &     0.35 &     0.01 &     0.01 &     0.30 &         \\
\bf{ 8228742} &   1171.0 &   1375.9 &   1036.9 &      --- &   62.0 &    4.8 &   6042 &   6048 &   6096 &   6093 &   6180 &      --- &     1.37 &     1.41 &     1.31 &     1.87 &     1.89 &     1.84 &     4.03 &     4.03 &     4.02 & 2,13,17 \\
              &     34.0 &     13.8 &     10.4 &      --- &    0.6 &    --- &     60 &     56 &    206 &    108 &    124 &      --- &     0.19 &     0.19 &     0.01 &     0.10 &     0.10 &     0.01 &     0.01 &     0.01 &     0.03 &   37    \\
\bf{ 8379927} &   2669.0 &   2981.4 &   2364.9 &      --- &  120.0 &   10.6 &   5998 &   5939 &   6096 &   5975 &   6107 &      --- &     1.15 &     1.17 &     1.09 &     1.13 &     1.14 &     1.11 &     4.39 &     4.39 &     4.25 & 2,3,37  \\
              &     80.1 &     29.8 &     23.6 &      --- &    1.0 &    --- &    108 &      --- &      --- &    158 &    164 &      --- &     0.17 &     0.17 &     0.03 &     0.06 &     0.06 &     0.02 &     0.01 &     0.01 &     0.21 &   40    \\
\bf{ 8394589} &   2165.0 &   2547.2 &   1891.5 &      --- &  109.5 &    8.1 &   6111 &   6105 &   5974 &   6120 &   6114 &      --- &     0.90 &     0.93 &     0.94 &     1.11 &     1.13 &     1.12 &     4.30 &     4.30 &     3.98 & 2,17,39 \\
              &    124.0 &     25.5 &     18.9 &      --- &    1.9 &    --- &    116 &     74 &    260 &    181 &    245 &      --- &     0.24 &     0.24 &     0.04 &     0.11 &     0.11 &     0.02 &     0.02 &     0.02 &     0.21 &   40    \\
\bf{ 8524425} &   1081.0 &   1128.0 &    831.3 &      --- &   59.4 &    5.0 &   5634 &   5513 &   5515 &   5541 &   5651 &      --- &     1.19 &     1.18 &     1.00 &     1.84 &     1.83 &     1.73 &     3.98 &     3.98 &     3.98 & 2,13,17 \\
              &     28.0 &     11.3 &      8.3 &      --- &    0.6 &    --- &     60 &     57 &    188 &    187 &    171 &      --- &     0.16 &     0.16 &     0.07 &     0.09 &     0.09 &     0.02 &     0.01 &     0.01 &     0.03 &   39    \\
\bf{ 8561221} &    491.0 &    488.0 &    370.5 &      --- &   29.8 &    2.4 &   5245 &   5183 &   5374 &   5238 &   5705 &      --- &     1.56 &     1.56 &     1.55 &     3.19 &     3.19 &     3.18 &     3.62 &     3.62 &     3.61 & 13,25   \\
              &      5.0 &      4.9 &      3.7 &      --- &    0.1 &    --- &     60 &     62 &    199 &    118 &     79 &      --- &     0.10 &     0.10 &     0.13 &     0.08 &     0.08 &     0.18 &     0.01 &     0.01 &     0.03 &         \\
\bf{ 8694723} &   1384.0 &   1661.8 &   1261.7 &      --- &   74.9 &    5.4 &   6258 &   6230 &   6355 &   6149 &   6272 &      --- &     1.08 &     1.14 &     0.96 &     1.51 &     1.56 &     1.44 &     4.11 &     4.11 &     3.97 & 2,3,39  \\
              &     41.5 &     16.6 &     12.6 &      --- &    0.8 &    --- &    117 &     36 &    137 &     96 &    109 &      --- &     0.17 &     0.17 &     0.03 &     0.09 &     0.09 &     0.02 &     0.01 &     0.01 &     0.21 &   40    \\
\bf{ 8702606} &    664.0 &    688.7 &    554.5 &      --- &   39.7 &    3.5 &   5540 &   5396 &   5445 &   5489 &   6057 &      --- &     1.35 &     1.34 &     1.27 &     2.51 &     2.50 &      --- &     3.77 &     3.77 &     3.76 & 2,13,22 \\
              &     16.0 &      6.9 &      5.5 &      --- &    0.5 &    --- &     60 &     70 &    192 &    171 &    108 &      --- &     0.19 &     0.19 &      ---  &     0.14 &     0.14 &      --- &     0.01 &     0.01 &     0.03 &         \\
\bf{ 8760414} &   2384.0 &   2628.3 &   2041.6 &      --- &  117.1 &    5.6 &   5850 &   5925 &   5981 &   5873 &   6019 &      --- &     0.88 &     0.88 &     0.78 &     1.06 &     1.06 &     1.01 &     4.33 &     4.33 &     3.94 & 2,17,39 \\
              &    121.0 &     26.3 &     20.4 &      --- &    0.4 &    --- &    166 &     59 &    194 &    251 &    252 &      --- &     0.18 &     0.18 &     0.01 &     0.08 &     0.08 &     0.01 &     0.01 &     0.01 &     0.26 &   40    \\
\bf{ 9025370} &   2653.0 &   3246.7 &   2540.0 &      --- &  133.3 &    8.7 &   5704 &   5704 &   5630 &   6238 &   6381 &      --- &     0.70 &     0.70 &     0.84 &     0.90 &     0.90 &     0.96 &     4.37 &     4.37 &     --- & 2,17    \\
              &    215.0 &     32.5 &     25.4 &      --- &    1.9 &    --- &     99 &     71 &    116 &    183 &    222 &      --- &     0.23 &     0.23 &     0.13 &     0.11 &     0.11 &     0.05 &     0.01 &     0.01 &     --- &         \\
\bf{ 9098294} &   2233.0 &   2494.9 &   1949.7 &      --- &  108.8 &    5.9 &   5766 &   5756 &   5930 &   5938 &   6092 &      --- &     0.96 &     0.96 &     1.00 &     1.14 &     1.14 &     1.15 &     4.30 &     4.30 &     4.27 & 2,17,39 \\
              &     75.0 &     24.9 &     19.5 &      --- &    1.7 &    --- &     96 &     60 &    269 &    168 &    167 &      --- &     0.18 &     0.18 &     0.03 &     0.08 &     0.08 &     0.01 &     0.01 &     0.01 &     0.21 &   40    \\
\bf{ 9139151} &   2610.0 &   2972.7 &   2270.0 &      --- &  116.7 &   10.1 &   6125 &   6116 &   6265 &   6062 &   6022 &      --- &     1.22 &     1.26 &     1.14 &     1.18 &     1.20 &     1.15 &     4.38 &     4.38 &     4.38 & 2,13,39 \\
              &     78.3 &     29.7 &     22.7 &      --- &    2.1 &    --- &     60 &     53 &    214 &    138 &    180 &      --- &     0.22 &     0.22 &     0.03 &     0.08 &     0.08 &     0.01 &     0.02 &     0.02 &     0.03 &         \\
\bf{ 9139163} &   1608.0 &   2202.8 &   1619.8 &   1179.8 &   81.0 &    6.6 &   6400 &   6432 &   6395 &   6431 &   6484 &   6512 &     1.25 &     1.36 &     1.36 &     1.50 &     1.56 &     1.53 &     4.18 &     4.18 &     4.18 & 2,13,17 \\
              &     58.0 &     22.0 &     16.2 &     11.8 &    1.1 &    --- &     60 &     36 &    128 &     34 &     85 &    115 &     0.22 &     0.22 &     0.03 &     0.10 &     0.10 &     0.02 &     0.01 &     0.01 &     0.03 &   39    \\
\bf{ 9206432} &   1853.0 &   2289.7 &   1863.9 &   1355.8 &   84.3 &    7.7 &   6608 &   6524 &   6597 &   6306 &   6482 &   6433 &     1.63 &     1.86 &     1.40 &     1.58 &     1.69 &     1.48 &     4.25 &     4.25 &     4.23 & 2,13,17 \\
              &     46.0 &     22.9 &     18.6 &     13.6 &    1.3 &    --- &     60 &     56 &    174 &     63 &     71 &    110 &     0.24 &     0.24 &     0.03 &     0.10 &     0.10 &     0.01 &     0.01 &     0.01 &     0.03 &   39    \\
\bf{ 9410862} &   2261.0 &   2449.0 &   1912.1 &      --- &  107.0 &    7.8 &   6024 &   6024 &   6203 &   5799 &   5935 &      --- &     1.11 &     1.13 &     1.12 &     1.21 &     1.22 &     1.22 &     4.32 &     4.32 &     --- & 2,17    \\
              &     67.8 &     24.5 &     19.1 &      --- &    1.9 &    --- &    100 &    118 &      --- &    189 &    183 &      --- &     0.21 &     0.21 &     0.11 &     0.09 &     0.09 &     0.05 &     0.02 &     0.02 &     --- &         \\
\bf{ 9574283} &    455.0 &    459.3 &    370.6 &      --- &   29.9 &    3.0 &   5120 &      --- &      --- &   5328 &   6044 &      --- &     1.16 &     1.18 &     1.07 &     2.88 &     2.90 &      --- &     3.59 &     3.59 &     --- & 2,17,22 \\
              &     10.0 &      4.6 &      3.7 &      --- &    0.8 &    --- &     55 &      --- &      --- &    163 &     93 &      --- &     0.22 &     0.22 &      ---  &     0.23 &     0.23 &      --- &     0.02 &     0.02 &     --- &         \\
\bf{ 9812850} &   1195.0 &   1558.0 &   1171.2 &    877.2 &   65.1 &    4.1 &   6258 &   6272 &   6393 &   6336 &   6436 &   6557 &     1.22 &     1.29 &     1.39 &     1.73 &     1.78 &     1.75 &     4.05 &     4.05 &     3.94 & 2,17,39 \\
              &     60.0 &     15.6 &     11.7 &      8.8 &    1.1 &    --- &     97 &     51 &    223 &     78 &    118 &    126 &     0.29 &     0.29 &     0.05 &     0.16 &     0.16 &     0.02 &     0.02 &     0.02 &     0.21 &   40    \\
\bf{ 9955598} &   3546.0 &   3530.0 &   2995.3 &      --- &  153.0 &    9.5 &   5264 &   5355 &   5480 &   5236 &   5842 &      --- &     0.85 &     0.85 &     0.89 &     0.88 &     0.88 &     0.88 &     4.48 &     4.48 &     4.29 & 2,29,39 \\
              &    119.0 &     35.3 &     30.0 &      --- &    3.1 &    --- &     95 &     66 &    197 &    359 &    238 &      --- &     0.18 &     0.18 &     0.02 &     0.07 &     0.07 &     0.01 &     0.02 &     0.02 &     0.12 &   40    \\
\bf{10018963} &    987.0 &   1168.7 &    866.0 &      --- &   55.2 &    5.1 &   6145 &   6125 &   6285 &   6091 &   6175 &      --- &     1.32 &     1.37 &     1.18 &     1.99 &     2.03 &     1.92 &     3.96 &     3.96 &     3.95 & 2,17,39 \\
              &     32.0 &     11.7 &      8.7 &      --- &    0.5 &    --- &    112 &     43 &    131 &    111 &    128 &      --- &     0.21 &     0.21 &     0.03 &     0.12 &     0.12 &     0.02 &     0.01 &     0.01 &     0.21 &   40    \\
\bf{10162436} &    968.0 &   1370.4 &    977.0 &    671.1 &   55.5 &    3.6 &   6149 &   6155 &   6397 &   6427 &   6485 &   6505 &     1.22 &     1.27 &     1.23 &     1.93 &     1.97 &     1.90 &     3.95 &     3.95 &     3.95 & 2,17,39 \\
              &     49.0 &     13.7 &      9.8 &      6.7 &    0.7 &    --- &    115 &      --- &      --- &     41 &     98 &    136 &     0.28 &     0.28 &     0.02 &     0.16 &     0.16 &     0.02 &     0.01 &     0.01 &     0.21 &   40    \\
\bf{10355856} &   1330.0 &   1823.7 &   1308.8 &      --- &   68.1 &    4.7 &   6351 &   6326 &   6490 &   6427 &   6441 &      --- &     1.41 &     1.52 &     1.32 &     1.76 &     1.82 &     1.67 &     4.10 &     4.10 &     3.93 & 2,17,39 \\
              &     42.0 &     18.2 &     13.1 &      --- &    0.7 &    --- &    118 &     47 &    189 &     32 &     85 &      --- &     0.23 &     0.23 &     0.03 &     0.11 &     0.11 &     0.01 &     0.01 &     0.01 &     0.21 &   40    \\
\bf{10454113} &   2261.0 &   2594.3 &   2019.3 &      --- &  103.8 &    8.6 &   6120 &   6044 &   6179 &   6078 &   6148 &      --- &     1.27 &     1.31 &     1.19 &     1.29 &     1.31 &     1.25 &     4.32 &     4.32 &     4.31 & 2,13,17 \\
              &     62.0 &     25.9 &     20.2 &      --- &    1.3 &    --- &     60 &      --- &    144 &    122 &    143 &      --- &     0.19 &     0.19 &     0.04 &     0.07 &     0.07 &     0.01 &     0.01 &     0.01 &     0.03 &   39    \\
\bf{10516096} &   1700.0 &      --- &      --- &      --- &   84.6 &    --- &   5928 &   6006 &   5775 &      --- &      --- &      --- &     1.19 &     1.20 &     1.12 &     1.45 &     1.46 &     1.42 &     4.19 &     4.19 &     4.24 & 17,37,40\\
              &     30.0 &      --- &      --- &      --- &    1.1 &    --- &     95 &     67 &    188 &      --- &      --- &      --- &     0.15 &     0.15 &     0.03 &     0.07 &     0.07 &     0.03 &     0.01 &     0.01 &     0.21 &         \\
\bf{10644253} &   2819.0 &   3234.8 &   2623.1 &      --- &  123.2 &    9.8 &   6030 &   6046 &   5932 &   6103 &   6279 &      --- &     1.22 &     1.25 &     1.13 &     1.14 &     1.15 &     1.11 &     4.41 &     4.41 &     4.40 & 2,13,17 \\
              &    131.0 &     32.3 &     26.2 &      --- &    2.7 &    --- &     60 &     60 &    160 &    182 &    191 &      --- &     0.30 &     0.30 &     0.05 &     0.11 &     0.11 &     0.02 &     0.02 &     0.02 &     0.03 &   39    \\
\bf{10909629} &    839.0 &      --- &    843.7 &      --- &   49.6 &    3.0 &   6046 &   6046 &   6133 &      --- &   6479 &      --- &     1.23 &     1.26 &     1.36 &     2.09 &     2.11 &     2.17 &     3.89 &     3.89 &     --- & 2,17    \\
              &     37.0 &      --- &      8.4 &      --- &    1.0 &    --- &     99 &    137 &      --- &      --- &     86 &      --- &     0.29 &     0.29 &     0.11 &     0.19 &     0.19 &     0.07 &     0.02 &     0.02 &     --- &         \\
\bf{10920273} &   1024.0 &   1103.1 &    826.6 &      --- &   57.1 &    4.9 &   5710 &   6189 &   6725 &   5707 &   5882 &      --- &     1.20 &     1.20 &     1.00 &     1.90 &     1.89 &     1.78 &     3.96 &     3.96 &     4.15 & 14,18,23\\
              &     64.0 &     11.0 &      8.3 &      --- &    0.6 &    --- &     75 &    216 &      --- &    319 &    299 &      --- &     0.30 &     0.30 &     0.04 &     0.17 &     0.17 &     0.02 &     0.01 &     0.01 &     0.08 &         \\
\bf{10963065} &   2184.0 &   2497.9 &   1885.3 &      --- &  102.6 &    7.3 &   6097 &   6116 &   6177 &   6054 &   6023 &      --- &     1.19 &     1.23 &     1.05 &     1.27 &     1.29 &     1.21 &     4.30 &     4.30 &     4.00 & 2,17,39 \\
              &     62.0 &     25.0 &     18.9 &      --- &    1.0 &    --- &    130 &     35 &    143 &    128 &    163 &      --- &     0.19 &     0.19 &     0.02 &     0.07 &     0.07 &     0.01 &     0.01 &     0.01 &     0.21 &   40    \\
\bf{11026764} &    895.0 &    945.6 &    698.3 &      --- &   50.2 &    4.5 &   5682 &   5374 &   5727 &   5605 &   5746 &      --- &     1.33 &     1.32 &     1.27 &     2.14 &     2.13 &     2.11 &     3.90 &     3.90 &     3.88 & 2,13,17 \\
              &     29.0 &      9.5 &      7.0 &      --- &    0.6 &    --- &     60 &     34 &    155 &    205 &    188 &      --- &     0.21 &     0.21 &     0.06 &     0.13 &     0.13 &     0.03 &     0.01 &     0.01 &     0.03 &   39    \\
\bf{11081729} &   1990.0 &     ---  &   2391.0 &   1802.4 &   90.7 &    5.9 &   6630 &   6534 &   6637 &   ---  &   6661 &   6738 &     1.51 &     1.73 &     1.26 &     1.47 &     1.57 &     1.38 &     4.28 &     4.28 &     4.25 & 2,13,17 \\
              &     84.0 &     24.9 &     18.5 &      --- &    1.4 &    --- &     60 &     55 &    207 &     84 &    166 &      --- &     0.31 &     0.31 &     0.03 &     0.11 &     0.11 &     0.02 &     0.01 &     0.01 &     0.03 &   39    \\
\bf{11244118} &   1420.0 &   1526.8 &   1169.9 &      --- &   71.3 &    5.5 &   5745 &   5729 &   5491 &   5736 &   5868 &      --- &     1.33 &     1.32 &     1.10 &     1.69 &     1.69 &     1.59 &     4.10 &     4.10 &     4.09 & 2,13,17 \\
              &     31.0 &     15.3 &     11.7 &      --- &    0.9 &    --- &     60 &     58 &    197 &    152 &    143 &      --- &     0.18 &     0.18 &     0.05 &     0.09 &     0.09 &     0.03 &     0.01 &     0.01 &     0.03 &   39    \\
\bf{11253226} &   1638.0 &   2150.4 &   1684.6 &   1194.6 &   76.9 &    4.4 &   6410 &   6572 &   6768 &   6402 &   6520 &   6451 &     1.63 &     1.77 &     1.41 &     1.70 &     1.77 &     1.55 &     4.19 &     4.19 &     3.96 & 2,17,39 \\
              &     48.0 &     21.5 &     16.8 &     11.9 &    1.0 &    --- &    125 &     38 &    156 &     41 &     68 &    118 &     0.28 &     0.28 &     0.05 &     0.11 &     0.11 &     0.02 &     0.01 &     0.01 &     0.21 &   40    \\
\bf{11295426} &   2154.0 &   2233.4 &   1766.4 &      --- &  101.2 &    5.8 &   5793 &   5838 &   5712 &   5539 &   5773 &      --- &     1.16 &     1.15 &     1.08 &     1.28 &     1.28 &     1.24 &     4.29 &     4.29 &     4.28 & 27,29,46\\
              &     13.0 &     22.3 &     17.7 &      --- &    1.0 &    --- &     74 &     87 &    235 &     99 &     78 &      --- &     0.09 &     0.09 &     0.05 &     0.04 &     0.04 &     0.02 &     0.01 &     0.01 &     0.06 &         \\
\bf{11395018} &    834.0 &    875.3 &    685.2 &      --- &   47.3 &    4.2 &   5445 &   5517 &   5458 &   5566 &   5965 &      --- &     1.29 &     1.28 &     1.27 &     2.20 &     2.19 &     2.18 &     3.86 &     3.86 &     3.84 & 2,18,23 \\
              &     50.0 &      8.8 &      6.9 &      --- &    0.5 &    --- &     85 &    101 &    376 &    345 &    264 &      --- &     0.31 &     0.31 &     0.04 &     0.19 &     0.19 &     0.02 &     0.01 &     0.01 &     0.12 &         \\
\bf{11414712} &    707.0 &    781.2 &    586.9 &      --- &   43.9 &    4.1 &   5635 &   5581 &   5563 &   5783 &   6063 &      --- &     1.11 &     1.10 &     1.26 &     2.20 &     2.20 &     2.34 &     3.80 &     3.80 &     3.80 & 2,13,17 \\
              &     20.0 &      7.8 &      5.9 &      --- &    0.7 &    --- &     60 &     33 &     81 &    141 &    117 &      --- &     0.18 &     0.18 &     0.08 &     0.14 &     0.14 &     0.06 &     0.02 &     0.02 &     0.03 &         \\
\bf{11713510} &   1241.0 &      --- &      --- &      --- &   68.9 &    --- &   5893 &   5893 &   6055 &      --- &      --- &      --- &     1.05 &     1.05 &     1.00 &     1.60 &     1.60 &     1.57 &     4.05 &     4.05 &     --- & 17,37   \\
              &     33.0 &      --- &      --- &      --- &    0.9 &    --- &      9 &    133 &      --- &      --- &      --- &      --- &     0.14 &     0.14 &     0.01 &     0.09 &     0.09 &     0.01 &     0.01 &     0.01 &     --- &         \\
\bf{11717120} &    585.0 &    583.7 &    434.3 &      --- &   37.8 &    4.2 &   5150 &   5034 &   5165 &   5263 &   5722 &      --- &     0.99 &     1.00 &     1.01 &     2.33 &     2.35 &     2.38 &     3.70 &     3.70 &     3.68 & 2,13,17 \\
              &      8.0 &      5.8 &      4.3 &      --- &    0.9 &    --- &     60 &     42 &    140 &    128 &     87 &      --- &     0.15 &     0.15 &     0.10 &     0.16 &     0.16 &     0.11 &     0.02 &     0.02 &     0.03 &         \\
\bf{11771760} &    535.0 &    652.7 &    477.2 &      --- &   32.2 &    3.1 &   6142 &      --- &      --- &   6142 &   6246 &      --- &     1.81 &     1.87 &     1.55 &     3.16 &     3.22 &     3.00 &     3.70 &     3.70 &     --- & 2,17    \\
              &     19.0 &      6.5 &      4.8 &      --- &    0.7 &    --- &     99 &      --- &      --- &    100 &    116 &      --- &     0.39 &     0.39 &     0.14 &     0.27 &     0.27 &     0.01 &     0.02 &     0.02 &     --- &         \\
\bf{11772920} &   3439.0 &   3709.5 &      --- &      --- &  157.4 &    8.7 &   5209 &   5371 &   5153 &   5790 &      --- &      --- &     0.68 &     0.68 &      ---  &     0.80 &     0.80 &      --- &     4.47 &     4.47 &     4.34 & 2,40    \\
              &    103.2 &     37.1 &      --- &      --- &    1.6 &    --- &     51 &     57 &    173 &    197 &      --- &      --- &     0.10 &     0.10 &      ---  &     0.04 &     0.04 &      --- &     0.01 &     0.01 &     0.23 &         \\
\bf{11807274} &   1496.0 &   1678.7 &   1355.2 &      --- &   75.1 &    5.7 &   6225 &   6107 &   6365 &   5943 &   6232 &      --- &     1.35 &     1.42 &     1.26 &     1.63 &     1.67 &     1.58 &     4.14 &     4.14 &     4.13 & 16,29   \\
              &     56.0 &     16.8 &     13.6 &      --- &    0.8 &    --- &     66 &     97 &    478 &    176 &    149 &      --- &     0.23 &     0.23 &     0.07 &     0.10 &     0.10 &     0.03 &     0.01 &     0.01 &     0.01 &         \\
\bf{12009504} &   1768.0 &   2003.2 &   1558.7 &      --- &   88.1 &    6.0 &   6099 &   6232 &   6022 &   5988 &   6139 &      --- &     1.16 &     1.20 &     1.12 &     1.40 &     1.42 &     1.38 &     4.21 &     4.21 &     4.00 & 2,17,39 \\
              &     40.0 &     20.0 &     15.6 &      --- &    1.2 &    --- &    125 &     61 &    168 &    115 &    115 &      --- &     0.18 &     0.18 &     0.02 &     0.08 &     0.08 &     0.04 &     0.01 &     0.01 &     0.21 &   40    \\
\bf{12069424} &   2101.0 &   2317.2 &   1802.2 &      --- &  103.4 &    5.8 &   5813 &   5790 &   5741 &   5874 &   6027 &      --- &     0.99 &     0.99 &     1.11 &     1.20 &     1.20 &     1.24 &     4.28 &     4.28 &     4.28 & 35,38,45\\
              &     63.0 &     23.2 &     18.0 &      --- &    1.0 &    --- &     18 &      --- &      --- &    166 &    162 &      --- &     0.13 &     0.13 &     0.02 &     0.06 &     0.06 &     0.01 &     0.01 &     0.01 &     0.02 &         \\
\bf{12069449} &   2552.0 &   2626.2 &   2114.5 &      --- &  116.7 &    6.6 &   5749 &   5745 &   5671 &   5493 &   5798 &      --- &     1.08 &     1.07 &     1.07 &     1.14 &     1.13 &     1.13 &     4.36 &     4.36 &     4.33 & 35,38,45\\
              &     76.6 &     26.3 &     21.1 &      --- &    1.2 &    --- &     17 &      --- &      --- &    260 &    212 &      --- &     0.14 &     0.14 &     0.02 &     0.06 &     0.06 &     0.01 &     0.01 &     0.01 &     0.02 &         \\
\bf{12258514} &   1440.0 &   1667.1 &   1251.8 &      --- &   74.5 &    4.9 &   5990 &   6017 &   6062 &   6052 &   6108 &      --- &     1.21 &     1.23 &     1.20 &     1.59 &     1.60 &     1.59 &     4.12 &     4.12 &     4.11 & 2,13,17 \\
              &     43.0 &     16.7 &     12.5 &      --- &    0.8 &    --- &     60 &      --- &      --- &    121 &    141 &      --- &     0.18 &     0.18 &     0.08 &     0.09 &     0.09 &     0.04 &     0.01 &     0.01 &     0.03 &   39    \\
\bf{12317678} &   1238.0 &   1681.9 &   1349.5 &    945.8 &   63.3 &    3.5 &   6401 &   6401 &   6760 &   6418 &   6588 &   6583 &     1.53 &     1.66 &     1.41 &     1.89 &     1.97 &     1.85 &     4.07 &     4.07 &     --- & 2,17    \\
              &     40.0 &     16.8 &     13.5 &      9.5 &    0.8 &    --- &     99 &     39 &    153 &     36 &     48 &     87 &     0.26 &     0.26 &     0.13 &     0.12 &     0.12 &     0.06 &     0.01 &     0.01 &     --- &         \\
\bf{12508433} &    793.0 &    786.7 &    650.6 &      --- &   44.9 &    3.8 &   5134 &   5161 &   5062 &   5223 &   5959 &      --- &     1.22 &     1.24 &     1.17 &     2.23 &     2.25 &     2.20 &     3.83 &     3.83 &     3.50 & 2,17,40 \\
              &     26.0 &      7.9 &      6.5 &      --- &    0.7 &    --- &    121 &     84 &    174 &    271 &    160 &      --- &     0.24 &     0.24 &     0.12 &     0.17 &     0.17 &     0.09 &     0.01 &     0.01 &     0.28 &         \\
\bf{    2151} &   1000.0 &   1108.0 &    831.9 &      --- &   57.6 &    5.1 &   5790 &   5955 &   5762 &   5826 &   6021 &      --- &     1.10 &     1.10 &     1.08 &     1.83 &     1.83 &     1.81 &     3.95 &     3.95 &     3.84 & 8,11,12 \\
              &     30.0 &     11.1 &      8.3 &      --- &    0.6 &    --- &     40 &      --- &      --- &    151 &    138 &      --- &     0.15 &     0.15 &     0.03 &     0.10 &     0.10 &     0.02 &     0.01 &     0.01 &     0.08 &         \\
\bf{   43587} &   2247.0 &   2485.3 &   1965.2 &      --- &  106.4 &    6.0 &   5947 &   5835 &   5897 &   5905 &   6082 &      --- &     1.10 &     1.12 &     1.04 &     1.21 &     1.22 &     1.19 &     4.31 &     4.31 &     4.37 & 10,41   \\
              &     15.0 &     24.9 &     19.7 &      --- &    1.1 &    --- &     17 &     79 &     64 &     71 &     63 &      --- &     0.07 &     0.07 &     0.01 &     0.03 &     0.03 &     0.04 &     0.01 &     0.01 &     0.04 &         \\
\bf{   49385} &   1013.0 &   1154.8 &    881.3 &      --- &   56.3 &    4.1 &   6095 &      --- &   6241 &   5960 &   6146 &      --- &     1.31 &     1.35 &     1.25 &     1.96 &     1.99 &     1.94 &     3.97 &     3.97 &     4.00 & 19,20   \\
              &      3.0 &     11.5 &      8.8 &      --- &    0.6 &    --- &     65 &      --- &      --- &     45 &     38 &      --- &     0.09 &     0.09 &     0.05 &     0.06 &     0.06 &     0.03 &     0.01 &     0.01 &     0.06 &         \\
\bf{   49933} &   1760.0 &      --- &   2053.9 &   1670.8 &   86.1 &    2.2 &   6522 &      --- &   6922 &      --- &   6643 &   6768 &     1.29 &     1.43 &     1.28 &     1.45 &     1.53 &     1.46 &     4.23 &     4.23 &     4.00 & 1,34,47 \\
              &     52.8 &      --- &     20.5 &     16.7 &    0.9 &    --- &     38 &      --- &      --- &      --- &     27 &     29 &     0.18 &     0.18 &     0.01 &     0.08 &     0.08 &     0.01 &     0.01 &     0.01 &     0.06 &         \\
\bf{   52265} &   2090.0 &   2389.8 &   1808.0 &      --- &   98.1 &    8.2 &   6116 &   6096 &   6208 &   6054 &   6031 &      --- &     1.25 &     1.29 &     1.24 &     1.33 &     1.35 &     1.33 &     4.29 &     4.29 &     4.32 & 5,24,31 \\
              &     20.0 &     23.9 &     18.1 &      --- &    1.0 &    --- &    110 &      --- &      --- &     67 &     80 &      --- &     0.12 &     0.12 &     0.02 &     0.05 &     0.05 &     0.02 &     0.01 &     0.01 &     0.20 &         \\
\bf{  146233} &   3100.0 &   3202.1 &   2668.7 &      --- &  133.4 &   10.5 &   5693 &   6146 &   5680 &   5533 &   5943 &      --- &     1.12 &     1.11 &     1.03 &     1.05 &     1.05 &     1.02 &     4.44 &     4.44 &     4.48 & 7,32,47 \\
              &     93.0 &     32.0 &     26.7 &      --- &    1.3 &    --- &    108 &      --- &      --- &    267 &    201 &      --- &     0.18 &     0.18 &     0.01 &     0.06 &     0.06 &     0.01 &     0.01 &     0.01 &     0.06 &         \\
\bf{  181420} &   1610.0 &      --- &   1674.1 &      --- &   75.6 &    6.7 &   6580 &   6529 &   6590 &      --- &   6537 &      --- &     1.65 &     1.87 &     1.30 &     1.71 &     1.82 &     1.61 &     4.19 &     4.19 &     4.09 & 6,28,44 \\
              &     10.0 &      --- &     16.7 &      --- &    0.8 &    --- &    100 &     53 &    109 &      --- &     26 &      --- &     0.13 &     0.13 &     0.17 &     0.06 &     0.06 &     0.10 &     0.01 &     0.01 &     0.15 &         \\
\bf{  181907} &     28.5 &     24.2 &      --- &      --- &    3.5 &    0.7 &   4725 &   4744 &   4758 &   4741 &      --- &      --- &     1.30 &     1.41 &     1.43 &    12.42 &    12.91 &    12.93 &     2.37 &     2.36 &     2.35 & 15,26,30\\
              &      0.7 &      0.2 &      --- &      --- &    0.1 &    --- &     65 &      --- &      --- &    132 &      --- &      --- &     0.20 &     0.20 &     0.23 &     0.76 &     0.76 &     0.44 &     0.01 &     0.01 &     0.04 &         \\
\bf{  203608} &   2600.0 &      --- &   2488.0 &      --- &  120.3 &    6.7 &   6253 &   6165 &   6266 &      --- &   6371 &      --- &     1.08 &     1.14 &     0.93 &     1.10 &     1.13 &     1.06 &     4.39 &     4.39 &     4.36 & 42,43   \\
              &      0.5 &      --- &     24.9 &      --- &    1.2 &    --- &     32 &      --- &      --- &      --- &     26 &      --- &     0.05 &     0.05 &     0.01 &     0.03 &     0.03 &     0.01 &     0.01 &     0.01 &     0.01 &         \\
\bf{ Procyon A} &   1014.0 &      --- &   1129.7 &    739.4 &   55.2 &    2.5 &   6530 &   6544 &   6633 &      --- &   6602 &   6546 &     1.46 &     1.63 &     1.48 &     2.03 &     2.15 &     2.03 &     3.99 &     3.99 &     4.05 & 48,49,50\\
              &     11.0 &      --- &     11.3 &      7.4 &    0.5 &    --- &     90 &      --- &      --- &      --- &     20 &     41 &     0.13 &     0.13 &     0.01 &     0.07 &     0.07 &     0.01 &     0.01 &     0.01 &     0.04 & 51,52,53\\
\bf{$\bigodot$} &   3050.0 &   3256.6 &   2555.2 &   1879.5 &  136.0 &    9.8 &   5777 &      --- &      --- &   5742 &   5831 &   5840 &     1.00 &     1.00 &     1.00 &     1.00 &     1.00 &     1.00 &     4.44 &     4.44 &     4.44 &         \\
              &     49.9 &     32.6 &     25.6 &     18.8 &    0.1 &    --- &     20 &      --- &      --- &    141 &    137 &    148 &     0.06 &     0.06 &     0.01 &     0.02 &     0.02 &     0.01 &     0.01 &     0.01 &     0.01 &         \\
\hline
\end{longtable}
~~\\
Ref.  $-$ 1: Appourchaux et al.(2008), 2: Appourchaux et al.(2012), 3: Appourchaux et al.(2014), 4: Ballard et al.(2014), 5: Ballot et al.(2011), 6: Barban et al.(2009), 7: Bazot et al.(2012), 8: Bedding et al.(2007), 9: Benomar et al.(2014), 10: Boumier et al.(2014), 11: Brand{\~{a}}o et al.(2011), 12: Bruntt et al.(2010), 13: Bruntt et al.(2012), 14: Campante et al.(2011), 15: Carrier et al.(2010), 16: Chaplin et al.(2013), 17: Chaplin et al.(2014), 18: Creevey et al.(2012), 19: Deheuvels et al.(2010), 20: Deheuvels \& Michel(2011), 21: Deheuvels et al.(2012), 22: Deheuvels et al.(2014), 23: Do{\u{g}}an et al.(2013), 24: Escobar et al.(2012), 25: Garc{\'{i}}a et al.(2014), 26: Ghezzi \& Johnson(2015), 27: Gilliland et al.(2013), 28: Hekker \& Ball(2014), 29: Huber et al.(2013), 30: Lagarde et al.(2015), 31: Lebreton \& Goupil(2014), 32: Li et al.(2012), 33: Lillo-Box et al.(2014), 34: Liu et al.(2014), 35: Lund et al.(2014), 36: Marcy et al.(2014), 37: Mathur et al.(2012), 38: Metcalfe et al.(2012), 39: Metcalfe et al.(2014), 40: Molenda-{\.{Z}}akowicz et al.(2013), 41: Morel et al.(2013), 42: Mortier et al.(2014), 43: Mosser et al.(2008), 44: Ozel et al.(2013), 45: Ram{\'{i}}rez et al.(2011), 46: Santos et al.(2013), 47: Soubiran et al.(2010), 48: Aufdenberg, Ludwig  \&  Kervella (2005), 49: Bond et al. (2015),  50: Huber et al. (2011a), 51: Bedding et. al. (2010), 52: Eggenberger et al. (2004),  53: Fuhrmann et al. (1997)
\end{landscape}
\twocolumn
\label{lastpage}

\end{document}